\documentclass[11pt]{report}

\usepackage[margin=1in]{geometry}
\usepackage{setspace}
\usepackage[hyperfootnotes=false]{hyperref}
\hypersetup{ 		
    pdfauthor={Aidar Zhetessov},        
    colorlinks=true,			
    linkcolor=NavyBlue,
    urlcolor=NavyBlue,
    citecolor=NavyBlue		
}

\usepackage[pdftex]{graphicx}
\graphicspath{{./fig/}}
\usepackage[usenames,dvipsnames]{color}
\usepackage{epstopdf}
\usepackage{graphicx}
\usepackage[final]{pdfpages}
\usepackage{subcaption}
\usepackage[export]{adjustbox}
\usepackage{array}
\newcolumntype{P}[1]{>{\centering\arraybackslash}p{#1}}

\usepackage{fancyhdr}
\usepackage{amsmath,amsthm,amssymb,mathrsfs,dsfont,amsmath}
\newenvironment{rcases}
{\left.\begin{aligned}}
	{\end{aligned}\right\rbrace}

\newcommand*\mbf[1]{\ensuremath{\mathbf{#1}}}  
\newcommand*\dt[0]{\frac{d}{dt}} 

\newcommand\oprocendsymbol{\hbox{$\square$}}
\newcommand\oprocend{\relax\ifmmode\else\unskip\hfill\fi\oprocendsymbol}

\newcommand{\real}[0]{\mathbb R}

\newcommand{\Rtheta}{\mathbf{R}_{\theta_\mathsf{dq}} }
\newcommand{\Et}{\mathbf{E}^\top}
\newcommand{\E}{\mathbf{E}}
\newcommand{\Lt}{L_\mathsf{t}}
\newcommand{\Zt}{\mathbf{Z}_\mathsf{t}}
\newcommand{\Zs}{\mathbf{Z}_s}
\newcommand{\Yc}{\mathbf{Y}_c}
\newcommand{\is}{i_{s,\mathsf{dq}}}

\newcommand{\itdqdot}{\dot{i}_\mathsf{t,dq}}
\newcommand{\itdq}{i_\mathsf{t,dq}}
\newcommand{\Vdq}{v_\mathsf{dq}}

\begin{document}

\lhead[<even output>]{}
\chead[<even output>]{}
\rhead[<even output>]{\thepage}
\cfoot{} 

\title{Decentralized control methodology for multi-machine/multi-converter power systems}
\date{}
\begin{titlepage}

\newcommand{\HRule}{\rule{\linewidth}{0.5mm}} 



\begin{figure}
	\begin{subfigure}{0.5\textwidth}
		\includegraphics[width=8cm,left]{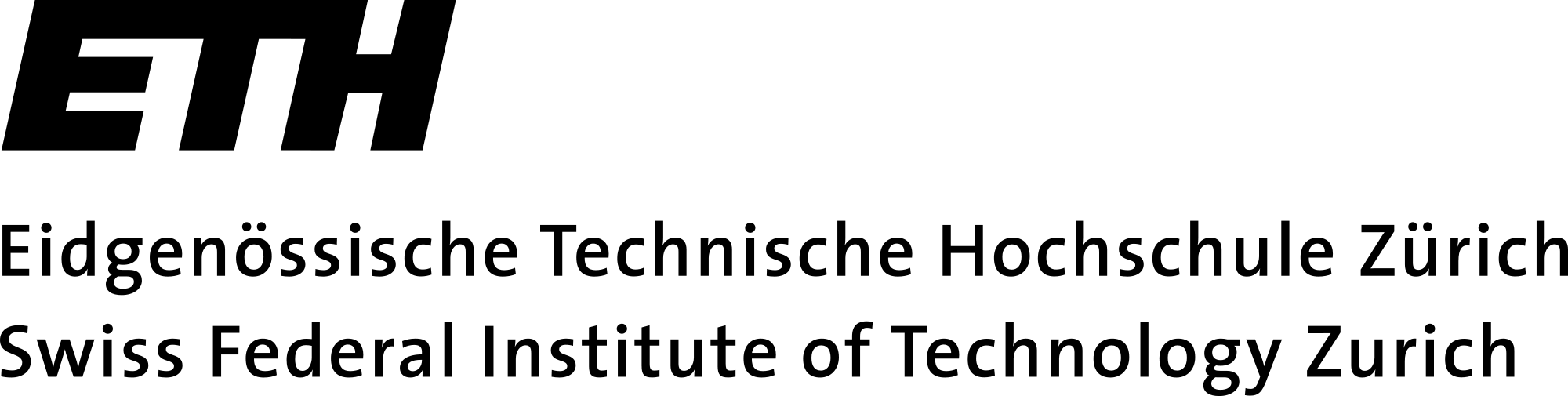}
	\end{subfigure}
	\begin{subfigure}{0.5\textwidth}
		\includegraphics[width=3.3cm,right]{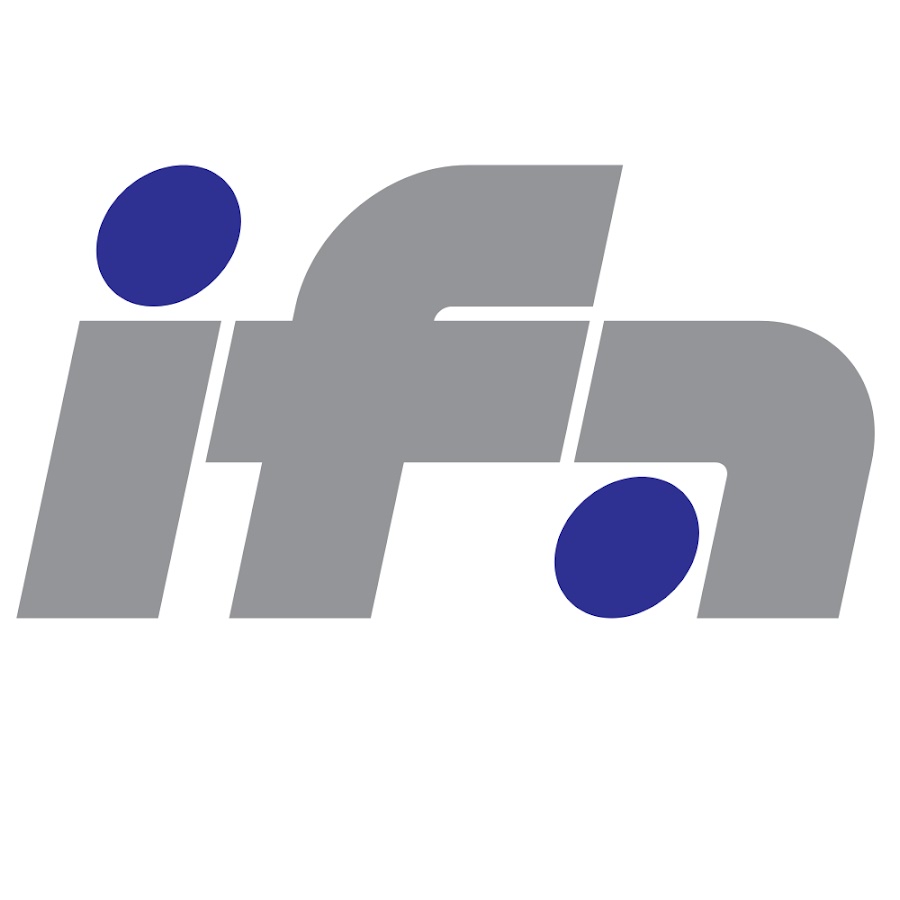}
	\end{subfigure}
\end{figure}
 

\center 

\makeatletter
\HRule \\[0.2cm]
{ \huge \bfseries \@title}\\[0.4cm] 

\includegraphics[width=0.75\textwidth,center]{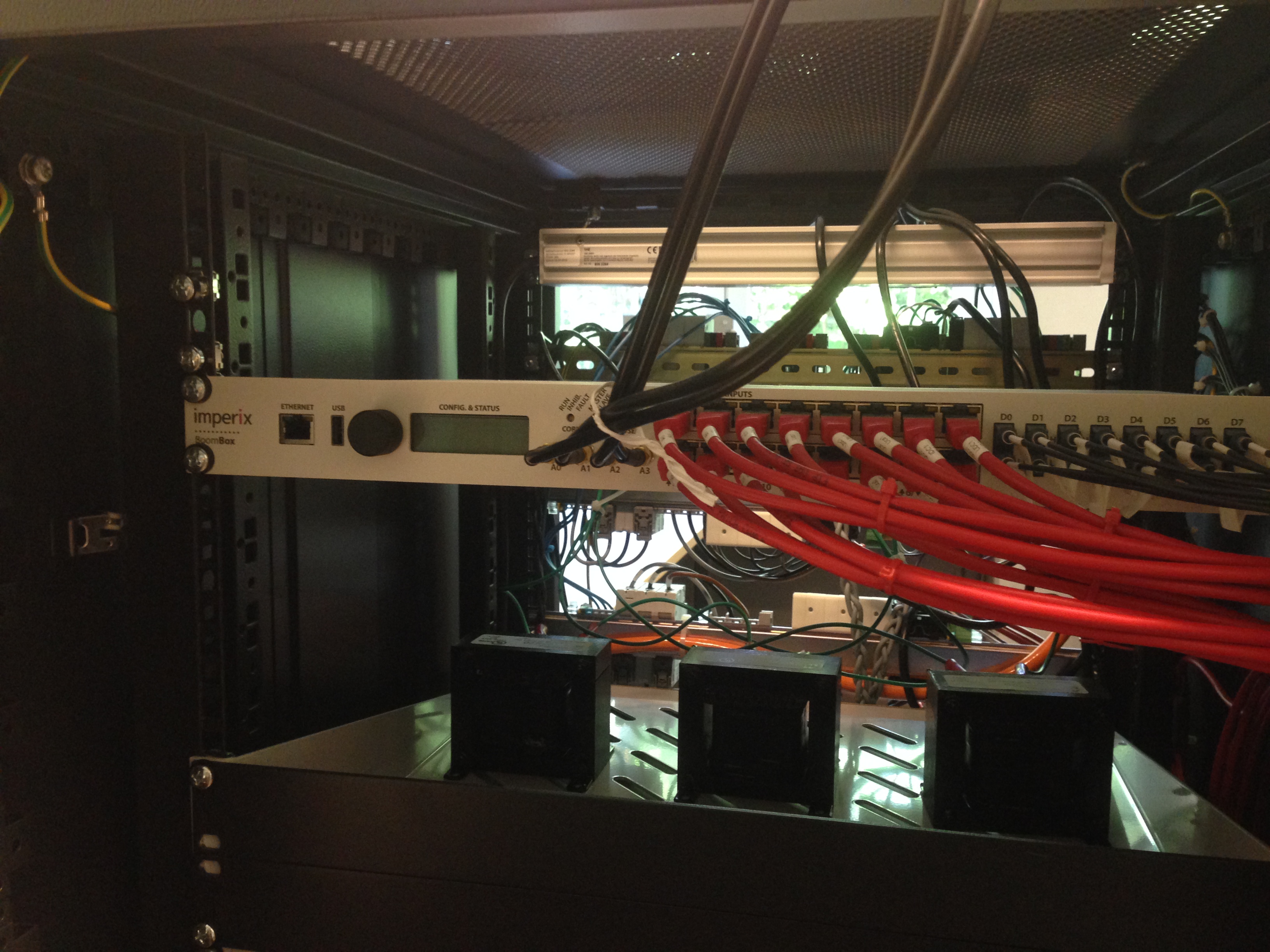}
\\[0.5cm]
\textsc{\LARGE Semester Thesis}\\[0.2cm] 
\textsc{\Large Spring Semester 2018}\\[0.2cm] 
\HRule \\[0.5cm]


\begin{minipage}{0.4\textwidth}
\begin{flushleft} \large
\emph{Author:}\\
Aidar Zhetessov \\
ID: 15-948-268
\end{flushleft}
\end{minipage}
~
\begin{minipage}{0.4\textwidth}
\begin{flushright} \large
\emph{Supervisor:} \\
Catalin Arghir \\[1.2em] 
\emph{Professor:} \\
Prof. Dr. F. D\"orfler 
\end{flushright}
\end{minipage}\\[1cm]
\makeatother



{\large June 4, 2018}\\[2cm] 

\vfill 

\end{titlepage}
\begin{abstract}
	In this project we evaluate a framework for synchronization of mixed machine-converter power grids. Synchronous machines are assumed to be actuated by mechanical torque injections, while the converters by DC-side current injections. As this approach is based on model-matching, the converter's modulation angle is driven by the DC-side voltage measurement, while its modulation amplitude is assigned analogously to the electrical machine's excitation current.
	
	In this way we provide extensions to the swing-equations model, retaining physical interpretation, and design controllers that achieve various objectives: frequency synchronization while stabilizing an angle configuration and a bus voltage magnitude prescribed by an optimal power flow (OPF) set-point. We further discuss decentralization issues related to clock drifts, loopy graphs, model reduction, energy function selection and characterizations of operating points. Finally, a numerical evaluation is based on experiments from three- and two- bus systems.
\end{abstract}
\tableofcontents
\listoffigures
\chapter{Introduction}

\section{Background and motivation}
Modern power system is undergoing significant changes with more renewable energy sources (RES) penetrating conventional grid through DC/AC converters, or simply inverters. In the light of this trend preserving voltage and frequency stability in the grid is becoming a more and more difficult task due to intermittent nature and inherent lack of rotational inertia in RES. On the other hand, interfacing power electronics provide much more flexibility in control that could be used to make up for mentioned RES-inherent deficiencies.

A broad spectrum of inverter control approaches has two main categories: grid-forming and grid-following \cite{Bevrani}. In the grid-following control schemes an inverter, assuming stiff grid, simply attaches to the network phase through PLL to inject power. Since a stable low-inertia grid cannot be formed only by grid-following inverters \cite{Jouini}, grid-forming control approaches, which provide network stability and ancillary services, should be investigated, especially for islanded micro-grids.

There are several grid-forming control approaches including, but not limited to droop, virtual synchronous machine (VSM), virtual oscillator control (VOC) and IoT/ICT based approaches. A brief, yet comprehensive overview of those with further references is provided in \cite{Jouini}-\cite{Tayyebi}. Another promising approach called matching control was recently presented in \cite{Jouini}. It utilizes the similarities between synchronous generator (SG) and inverter dynamics under certain modulation of the latter. This allows not merely emulating SG behavior, but exactly match it, so that there is no observable difference from grid side. This approach has several advantages such as elimination of costly AC side measurements, associated redundant computations and consequent actuation delays, incorporation of DC side to the dynamics, which ameliorates the demanding time-scale separation condition, ease of DC side measurement and remaining degree of freedom for ancillary services.

Within this local model-matching an overarching secondary control needs to be designed to achieve network level objectives like frequency synchronization, voltage control and optimal power flow (OPF) imposition. Energy-based (or passivity-based) framework, inspired by \cite{Schaft}-\cite{Manfredi} and similar works, was pursued in previous works of IfA lab of ETH Z\"urich, and this thesis continues progress in that direction.
\pagebreak

\section{Problem formulation}
With this being said, the problem at hand could be formulated as follows: first, given a prescribed OPF set-points for a network, within the local model-matching, what are the references for every local generator/inverter? In other words, given network-level objectives, what should every separate generator/inverter do locally? And second, supposing that there is a way of explicit network-level objective characterization through local references, how to stabilize the desired set-points in a topology-independent \cite{Bukhsh} and decentralized \cite{Johnson} manner? Research in these directions is the main focus of this thesis.

\section{Matching-control review}
For in-depth discussion on matching control the reader is referred to \cite{Jouini}. Here only a brief review will be added to remind the main aspects related the present work.

As it was mentioned, matching control builds on similarities between inverter and SG dynamics. Specifically it was observed that under certain inverter modulation strategy, in which the modulation reference depends on virtual angle, being the integral of DC-side voltage, inverter and SG dynamics can be exactly model-matched, thus eliminating the necessity of emulation, virtual modeling etc. Mentioned model-match becomes more evident if one writes the dynamic equations of both systems as well as associated figure:

\begin{figure}[h]
	\centering
	\includegraphics[width=0.87\textwidth]{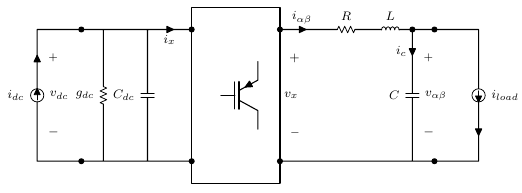}
	\caption{Circuit diagram of a 3-phase inverter \cite{Jouini}}
	\label{fig:jouini}
\end{figure}

\begin{equation}
\label{eq: matching-control}
\begin{split}
C_{dc} \dot{v}_{dc} &= -G_{dc} v_{dc} + i_{dc} - \frac{1}{2} i_{\alpha\beta}^\top m_{\alpha\beta}
\\
L\dot{i}_{\alpha\beta} &= -R i_{\alpha\beta} - v_{\alpha\beta} + \frac{1}{2} v_{dc}
\\
C \dot{v}_{\alpha\beta} &= - G v_{\alpha\beta} - i_l + i_{\alpha\beta}
\\
m_{\alpha\beta} &= \mu \begin{bmatrix} -sin(\theta_v) \\ cos(\theta_v) \end{bmatrix} \ \omega_v = \eta v_{dc}
\end{split}
\Leftrightarrow
\begin{split}
\dot \theta &= \omega
\\
M \dot{\omega} &= -D\omega + \tau_m + i_{\alpha\beta}^\top L_m i_f \begin{bmatrix}
-sin(\theta_v) \\ cos(\theta_v)
\end{bmatrix}
\\
L_s \dot{i}_{\alpha\beta} &= -R i_{\alpha\beta} - v_{\alpha\beta} - \omega_v L_m i_f \begin{bmatrix} -sin(\theta_v) \\ cos(\theta_v) \end{bmatrix}
\\
C \dot{v}_{\alpha\beta} &= - G v_{\alpha\beta} - i_l + i_{\alpha\beta}
\end{split}
\end{equation}

As it can be seen, inverter dynamics under mentioned modulation strategy match SG equations. This is the essence of local matching control. An advantage of proposed local control is in the model-equivalence of SGs and inverters, that somewhat simplifies the problem of overarching secondary control. The latter one, in turn, still needs to be designed for network-level objectives, such as load- and drift-robustness, voltage, frequency control etc.

\section{Modeling and notation}

Thanks to local matching, we have a unified framework for local modeling of inverters and SGs. Now, before tackling the questions stated in Problem formulation section, appropriate network model has to be defined. A good model should be simple enough, so that the problem is still tractable, and at the same time accurate enough, so that the main system aspects are captured.

\begin{figure}[h]
	\centering
	\includegraphics[width=0.8\textwidth]{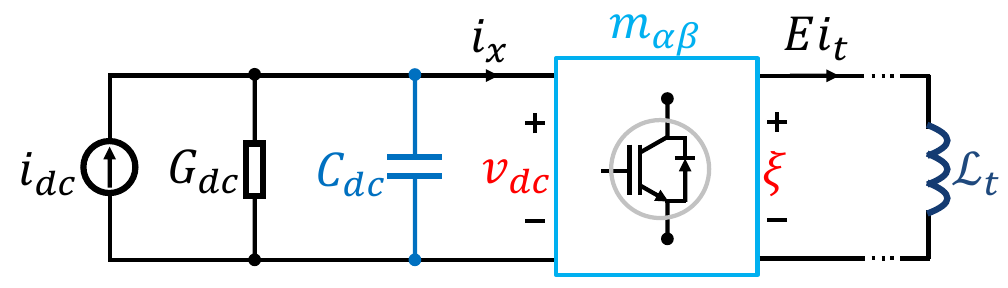}
	\caption{Matrix representation of reduced network model}
	\label{fig:reduced-model}
\end{figure}

Throughout the work two main models of the network were used - a reduced and full one. The adoption of two distinct models is the consequence of a good model definition above. Indeed, with full model all system aspects are captured, but analyzing it directly could become a difficult task. Therefore, as an intermediate step, reduced model - kind of a toy model, which preserves network, but simplifies node dynamics, could be used to gain some insight on the system. Matrix representation of a reduced model is depicted on Fig.\ref{fig:reduced-model}.

Here we assume that inverter output voltage vector $\xi$ consists of ideal sinusoidal voltages corresponding to inverters at every node and therefore there is no need in output $LC$ filter, so that resulting currents are directly injected into the network, represented as a Laplacian matrix. In SG notation reduced model neglects stator dynamics, assuming that generated $\xi$ is directly applied to the network. Proposed model, despite being a "toy", is quite general, since, in addition to preserving line dynamics, there are no constraints on network topology or, in other words, on graph incidence matrix $\E$.

\begin{figure}[h]
	\centering
	\includegraphics[width=\textwidth]{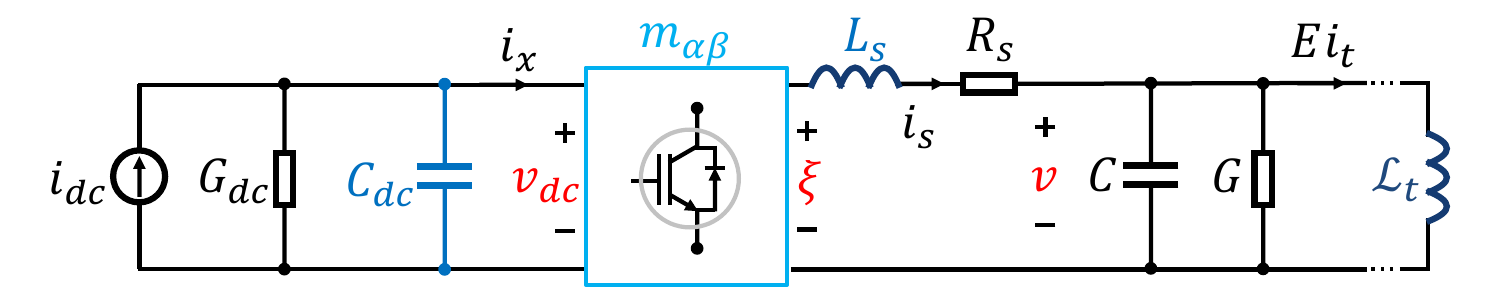}
	\caption{Matrix representation of full network model}
	\label{fig:full-model}
\end{figure}

Matrix representation of a full network model is shown on Fig.\ref{fig:full-model}. Here output filter dynamics and local load $G$ are added. In SG notation those correspond to stator and load dynamics. 

As one can see from Fig.\ref{fig:full-model}, there are several dynamic components at AC-side of the inverter/SG. A compact and comprehensive summary of whatever comes at AC-side at steady state for nominal frequency is given by so called $\Pi$-map:

\begin{equation}
\label{eq: Pi-map}
\Pi = \begin{bmatrix} \big(\mathbf{Z}_s + (\mathbf{Y}_c + \mathcal{L}_\mathsf{t})^{-1}\big)^{-1} \\ (\mathbf{Y}_c + \mathcal{L}_\mathsf{t})^{-1} \big(\mathbf{Z}_s + (\mathbf{Y}_c + \mathcal{L}_\mathsf{t})^{-1}\big)^{-1} \\ \mathbf{Z}_\mathsf{t} \Et (\mathbf{Y}_c + \mathcal{L}_\mathsf{t})^{-1} \big(\mathbf{Z}_s + (\mathbf{Y}_c + \mathcal{L}_\mathsf{t})^{-1}\big)^{-1} \end{bmatrix} = \begin{bmatrix}
\Pi_1 \\ \Pi_2 \\ \Pi_3
\end{bmatrix}
\end{equation}

Provided a fundamental component of SG/inverter output voltage vector $\xi$ at steady state, one can derive steady state fundamentals of stator currents, node voltages and transmission line currents throughout the network using this $\Pi$-map, comprising $\Pi_1,\Pi_2,\Pi_3$ respectively ($[\hat{i}_s^\top,\hat{v}^\top,\hat{i}_{\mathsf{t}}^\top]^\top = \Pi \xi$).

As it was already mentioned, this work, building on previous findings at IfA lab, continues to develop energy-based framework for decentralized secondary control in power systems. The notation was inherited from these previous works, so the last paragraph was taken as a reference to review the established notation that will be used in this work as well. For full derivation and description of notation the reader is referred to \cite{EnergySyn}.

Consider the $n$-dimensional angle variable $\theta \in \mathbb{T}^{n}$. Let $\Phi\in\real^{2n}$ be its Euclidean-space embedding such that $\Phi_i = \begin{bmatrix}\cos\theta_i \\ \sin\theta_i \end{bmatrix} \in \real^2$. Considering $I_n$ as the identity in $n$-dimensional Euclidean space, we denote $j =  \begin{bmatrix}0 & -1 \\ 1 & 0\end{bmatrix}$. Further define $\boldsymbol{j} = I_n \otimes j \in \real^{2n\times 2n}$, $\mathbf{R}_{\theta_i} = \begin{bmatrix} \Phi_i & j\Phi_i\end{bmatrix}\in\real^{2\times 2}$ and $\mathbf{R}_{\theta} = \text{blockdiag}\left\{\mathbf{R}_{\theta_i}\right\}\in \real^{2n\times 2n}$. We will aslo make use of $\mathbf{e}_1 =  \begin{bmatrix} 1 \\ 0 \end{bmatrix}$ and $\mathbf{e}_2 = j \mathbf{e}_1$. Transient error coordinates are introduced for all real-valued state-space variables, for example the following equation holds for $\forall t$: $v = \tilde{v} + \hat{v}$, where $\tilde{v}$ is a transient part and $\hat{v}$ is a steady state function of $\xi$ through $\Pi_2$. $\alpha\beta$-to-$\mathsf{dq}$ transformation is done using nominal frequency $\omega_0$ throughout the work.

\section{Outline}

The rest of report is organized as follows: Chapter 2 exploits the observation, found in \cite{GlobPVSyn}, to tackle the first question in problem formulation. Namely, it provides a bijective way of translation between network-level OPF-prescribed set-point and local controller references used in model-matching. Chapter 3 builds on obtained knowledge of local references and introduces the notion of communication graph to deal with circulating power flow problem within the topology-independence discussion. Inherent robustness issues of communication-based approach motivate the study of decentralization, a detailed discussion and substantial results on which are provided in Chapter 4. Finally, the summary of main findings and an outlook are given the concluding Chapter 5.
\chapter{Operating point characterization}

One of the main duties of central/emergency control level, provided grid state estimate, is to calculate the network-level OPF set-points. The OPF algorithm gives a consistent transmission line power flow references $P^*$ and $Q^*$ that preserve network frequency and bus voltage amplitudes within satisfactory limits \cite{Bevrani}. These power flow references are defined along transmission lines. The controllers, however, actuate on nodes of the network either in form of controllable DC current or mechanical torque injections. Therefore, to ensure these controllers have their references, a translation way from network-level OPF-prescribed set-points to local SG/inverter references should be established, especially if one considers further efforts in topology-independence and decentralization directions.

Both calculation and translation of OPF are difficult tasks, which even have their dedicated research directions. However the fact that OPF imposition applies to steady state significantly simplifies the translation task, because now a simpler controllers/models, that manage to impose an OPF at steady state and provide equations for bus voltages, can be studied. Taking mentioned control output bus voltages, one could reverse back into SG/inverter model at the nodes and derive local references ($I_r^*, \theta^*$ for SG. For inverter these parameters enter $m_{\alpha\beta}$ implicitly (\ref{eq: matching-control}))

One of such useful controllers was derived in \cite{GlobPVSyn}. In particular, it managed to almost globally stabilize the OPF-prescribed set-points on a network with shunt resistance (load), capacitance and controllable current source. Node schematic and proposed control law are provided below:

\begin{figure}[h]
	\centering
	\includegraphics[width=0.45\textwidth]{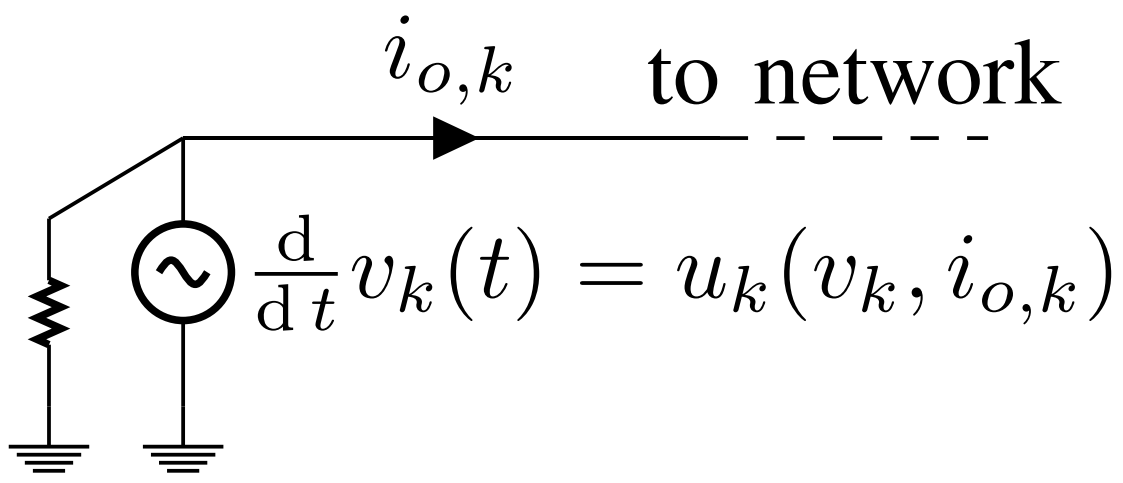}
	\caption{Schematics of decentralized control setup \cite{GlobPVSyn}}
	\label{fig:Colombino}
\end{figure}

\begin{equation}
\label{eq: control-Colombino}
\begin{split}
&\dt v_k = u_k(v_k,i_{o,k}) = \omega_0 j v_k + \eta \big(K_k v_k - R(\rho) i_{o,k}\big) + \alpha \Phi_k (v_k) v_k \ \ \text{where}\\
&\rho = tan^{-1} (\frac{\omega_0 l_{jk}}{r_{jk}}) \ \ K_k = \frac{1}{v_k^{*2}} R(\rho) \begin{bmatrix}
p_k^* & q_k^* \\ -q_k^* & p_k^*
\end{bmatrix} \ \ \Phi_k(v_k) = \frac{v_k^*-||v_k||}{v_k^*}
\end{split}
\end{equation}

What else was observed is that under $\frac{R}{L} = const.$ assumption for transmission lines, system dynamics can be represented current-free or as a function of node voltages only:

\begin{equation}
\label{eq: system-dynamics-voltages-only}
\begin{split}
&\dt v = \omega_0 J v + \eta \big(\mathcal{K}-\mathcal{L}\big) v + \alpha \Phi (v) v  \ \ \text{where}\\
&\mathcal{K} = diag\big(\{K_k\}_{k=1}^N\big) \ \ \Phi(v) = diag\big(\{\Phi_k(v_k)\}_{k=1}^N\big)
\end{split}
\end{equation}

If we slightly rewrite the equation (\ref{eq: system-dynamics-voltages-only}) we end up with the first vector field in equation (\ref{eq: vector-fields}), which is an explicit operating point characterization in terms of OPF. Obviously, vector field $f_1$ explicitly depends on power flow set-points $P^*$ and $Q^*$.

\begin{equation}
\label{eq: vector-fields}
\begin{split}
&\text{Explicit OPF (reference):} \ \dot{v}_\mathsf{dq} = f_1 (P^*,Q^*,||v_\mathsf{dq}^*||,v_\mathsf{dq}) \\
&\text{Implicit OPF (this work):} \ \dot{v}_\mathsf{dq} = f_2 (I_r^*,\theta_\mathsf{dq}^*,v_\mathsf{dq},...)
\end{split}
\end{equation}

Now, in this thesis we have an implicit operating point characterization in terms of OPF, since our vector field $f_2$ depends on local rotor current and angle references as well as some other parameters, which implicitly carry information on network-level OPF. Since the whole models in \cite{GlobPVSyn} (Fig.\ref{fig:Colombino}) and this thesis (Figures \ref{fig:reduced-model}-\ref{fig:full-model}) are very different, corresponding vector fields $f_1$ and $f_2$ are also very different, but what's in common is transmission line network, modeled in both works as an extended Laplacian matrix $\mathcal{L}$, or even a $\Pi$-configuration of transmission line in full model case. Indeed, reference \cite{GlobPVSyn} has controllable current sources attached to every node, which corresponds to the whole part on the left of bus voltage $v$ in Fig.\ref{fig:full-model}. The remaining shunt resistance, capacitance and network, is exactly the same as part of the full model on the right of bus voltage $v$ and is nothing more that $\Pi$-configuration of transmission lines. So, using this observation, one could argue that node (or bus) voltage set-points $v_{\alpha\beta}^*$ (or $v_\mathsf{dq}^*$) that impose OPF on a network in case of the first vector field at steady state, should also do the same in case of second vector field at steady state, since these set-points uniquely determine power flow in the network. Proposed argument can be used in both reduced and full models.

\section{Reduced model}

For a reduced model there is a direct relation between $\xi^*$ at steady state and $I_r^*,\theta^*$. Also recall the assumption of $\xi$ being an ideal sinusoid for a reduced model. So for local-reference-derivation, one could solve for node voltage set-points $v_{\alpha\beta}^*$ (or $v_\mathsf{dq}^*$) by solving steady state equation of vector field $f_1$ or, alternatively, by analyzing $ker(\mathcal{K}-\mathcal{L})$, provided $||v_\mathsf{dq}^*||$ holds (last part of controllers (\ref{eq: control-Colombino}) and (\ref{eq: system-dynamics-voltages-only}). Now, recall that $\xi_\mathsf{dq}^* = v_\mathsf{dq}^*$ for reduced model, moreover, $\xi_\mathsf{dq}^* = \mathbf{R}_{\theta_\mathsf{dq}^*} (L_m \otimes e_2) I_r^* \omega_0 \boldsymbol{1}$ \cite{EnergySyn}. From these one can derive $I_r^*,\theta_{\mathsf{dq}}^*$. The steps are summarized below:

\begin{equation}
\label{eq: OP-reduced-model}
\begin{split}
\begin{rcases}
&\text{Explicit OPF (reference):} \ \dot{v}_\mathsf{dq} = f_1 (P^*,Q^*,||v_\mathsf{dq}^*||,v_\mathsf{dq}) \\
&\text{Implicit OPF (this work):} \ \dot{v}_\mathsf{dq} = f_2 (I_r^*,\theta_\mathsf{dq}^*,v_\mathsf{dq},...)
\end{rcases}
&\text{at steady state} \ \ \dot{v}_\mathsf{dq} = 0 \rightarrow \\ 
&\rightarrow v_\mathsf{dq}^* \in ker(\mathcal{K}-\mathcal{L}) \\
&\rightarrow \xi_\mathsf{dq}^* = v_\mathsf{dq}^* \rightarrow I_r^*,\theta_\mathsf{dq}^*
\end{split}
\end{equation}

If we plug obtained local references $I_r^*,\theta_\mathsf{dq}^*$ into vector field $f_2$, we end up imposing very same OPF on our reduced model, provided that same network topology and parameters are used both in $f_1$ and $f_2$, including $\frac{R}{L} = const.$ assumption for transmission lines. Thus, there is a bijective way of translation between network-level OPF-prescribed set-points and local references for SG/inverter controllers of a reduced model. The argument is bidirectional, since provided consistent local references, it is possible (and even much easier) to derive network power flows.

\section{Full model}

For full model, the procedure has an additional step, as steady state node voltage set-points $v_\mathsf{dq}^*$ and SG/inverter fundamental of output voltage vector $\xi_\mathsf{dq}^*$ are related by $\Pi_2$-map ($\Pi_2 \xi_\mathsf{dq}^* = v_\mathsf{dq}^*$). Note that for full model assumption of noise-free $\xi_\mathsf{dq}^*$ does not hold, so $\Pi_2$ relates only fundamentals of $\xi_\mathsf{dq}^*$ and $v_\mathsf{dq}^*$. However, this is the actual reason of introducing $\Pi_2$, because in inverter notation $L_s, C$ (SG stator parameters) can be seen as output filter. Interestingly, it does not even need to be as large, since we assume simple 3-phase 2-level 6-switch inverter average model in local matching control. Provided high switching frequency for low-voltage (LV) range, high corner frequency of output $LC$ filter would suffice. Additional step for full model is included in (\ref{eq: OP-full-model}):

\begin{equation}
\label{eq: OP-full-model}
\begin{split}
\begin{rcases}
&\text{Explicit OPF (reference):} \ \dot{v}_\mathsf{dq} = f_1 (P^*,Q^*,||v_\mathsf{dq}^*||,v_\mathsf{dq}) \\
&\text{Implicit OPF (this work):} \ \dot{v}_\mathsf{dq} = f_2 (I_r^*,\theta_\mathsf{dq}^*,v_\mathsf{dq},...)
\end{rcases}
&\text{at steady state} \ \ \dot{v}_\mathsf{dq} = 0 \rightarrow \\ 
&\rightarrow v_\mathsf{dq}^* \in ker(\mathcal{K}-\mathcal{L}) \\
&\rightarrow \xi_\mathsf{dq}^*=\Pi_2^{-1} v_\mathsf{dq}^* \rightarrow I_r^*,\theta_\mathsf{dq}^*
\end{split}
\end{equation}

If we plug obtained local references $I_r^*,\theta_\mathsf{dq}^*$ into vector field $f_2$, we end up imposing very same OPF on our full model as well, provided that same network topology and parameters are used both in $f_1$ and $f_2$, including $\frac{R}{L} = const.$ assumption for transmission lines (same conditions). Thus, there is a bijective way of translation between network-level OPF-prescribed set-points and local references for SG/inverter controllers of a full model as well. This argument is also bidirectional, since provided consistent local references, it is possible (and even much easier) to derive network power flows. Mentioned bijective relation can be summarized in the following equation:

\begin{equation}
\label{eq: bijective-relation}
P^*,Q^*,||v_\mathsf{dq}^*|| \leftrightarrow I_r^*,\theta_{\mathsf{dq}}^*
\end{equation}

Existence of explicit translation (\ref{eq: bijective-relation}) for both models of this thesis allows studying decentralization and (to lesser extent) topology-independence, knowing that local references can impose global OPF on a network.

Note on angles: after estimating desired bus voltage angle at steady state for vector field $f_1$, \cite{GlobPVSyn} proposes selecting one node to be reference in terms of angle. Such node’s desired angle and initial condition are both zero. All others converge to the difference with respect to reference node angle. This is because ideally we do not care about exact phase of bus voltages. Power flow is determined by bus-voltage-differences and not by exact phase of a particular bus. Reduced and full models of this thesis are perfectly fine with mentioned angle treatment, since in our case we also care only about rotor/virtual angle differences.
\chapter{Topology independence}

\begin{figure}[h]
	\centering
	\begin{subfigure}{0.35\textwidth}
		\includegraphics[width=0.9\linewidth, height=5cm]{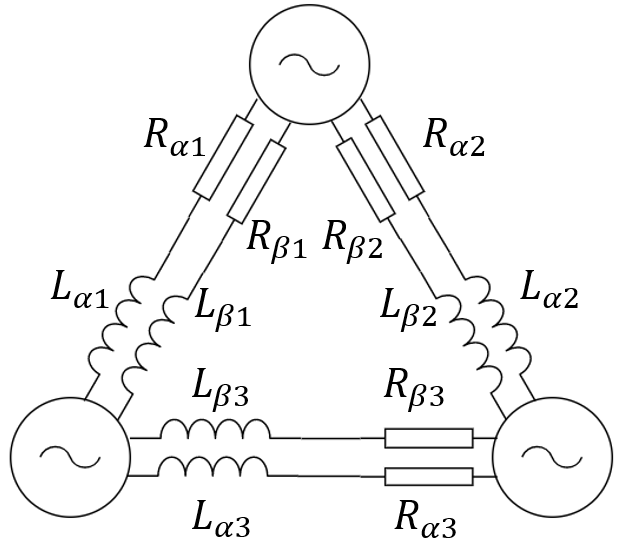}
		\caption{Cycled graph model}
		\label{fig:cycled-graph}
	\end{subfigure}
	\begin{subfigure}{0.6\textwidth}
		\includegraphics[width=1.02\linewidth, height=4cm]{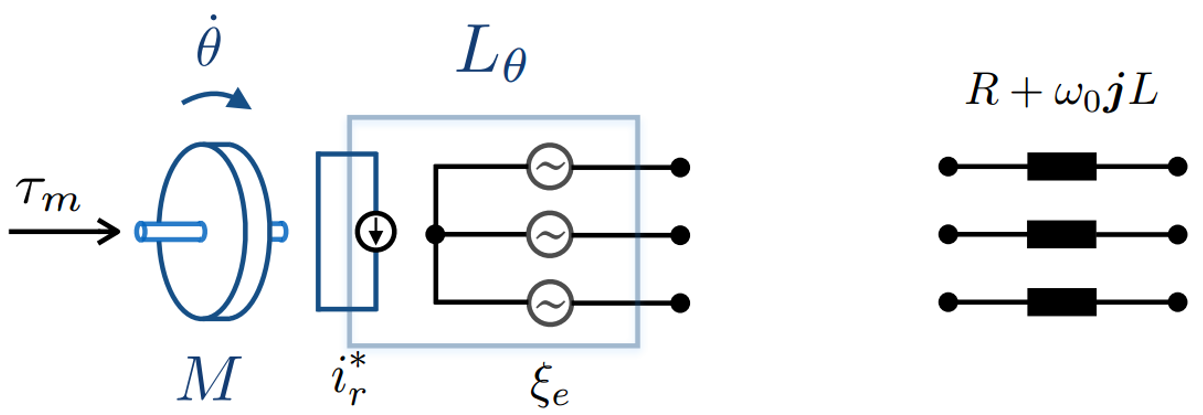}
		\caption{Single synchronous generator model}
		\label{fig:SG}
	\end{subfigure}
	\caption{Reduced power system model with cycled graph of generators}
	\label{fig:reduced-model-cycled}
\end{figure}

This chapter is dedicated to the consensus-oriented control of a large-cycled multi-machine power systems. Reduced model is to be used predominantly, since additional complication of stator in full model does not affect the control strategy. An example reduced model of a small, cycled power system is depicted in Fig.\ref{fig:reduced-model-cycled}.

As it can be seen from the figure, cycled graph comprises three nodes and edges connected in a triangle, which closes the cycle (loop). Graph edges model three-phase transmission lines in $\alpha\beta$-frame. A synchronous generator (SG), modeled and depicted in Fig.\ref{fig:SG}, is connected to every node. $\xi_e$ and $i_r^*$ denote line-connected SG EMF and reference rotor current.

When it comes to modeling networks, connecting ideal and fully controllable voltage/current sources to graph edges is practiced in some literature \cite{Bullo}-\cite{GlobPVSyn}. Similarly, when modeling SG, network is usually modeled as a simple infinite-bus \cite{Chapman}-\cite{Chiang}. Serving as a bridge, proposed reduced and full models have this additional advantage. On one hand, they do not oversimplify SG model, while on the other, preserve network dynamics.

First, the chapter begins with a motivation, introducing the problem of cycled topologies. Then it proposes the reduced model communication-based control implementation of identical/shifted rotor angle consensus with special focus on cycled graphs. Underlying remarks on robustness with respect to communication and clock drift are included in the end.

\section{Cycled topologies}

Several literature sources \cite{Bukhsh},\cite{Korsak}-\cite{DiffGeo} report the possibility as well as some underlying reasons of circulating power flow emergence in cycled topologies with five or more generators in the loop. Some conclusions are even backed up by real observations in power systems. It turns out that in large cycled topologies several local minima of network energy can arise. They correspond to circulating power flow in either clockwise or counter-clockwise direction. Interestingly, this power circulation is stable even in dissipative (resistive) transmission line networks under certain conditions. An example graph (or network), claimed to have real counterpart with observed circulation phenomenon, is shown in Fig.\ref{fig:Korsak}:

\begin{figure}[h]
	\centering
	\begin{subfigure}{0.45\textwidth}
		\includegraphics[width=\textwidth]{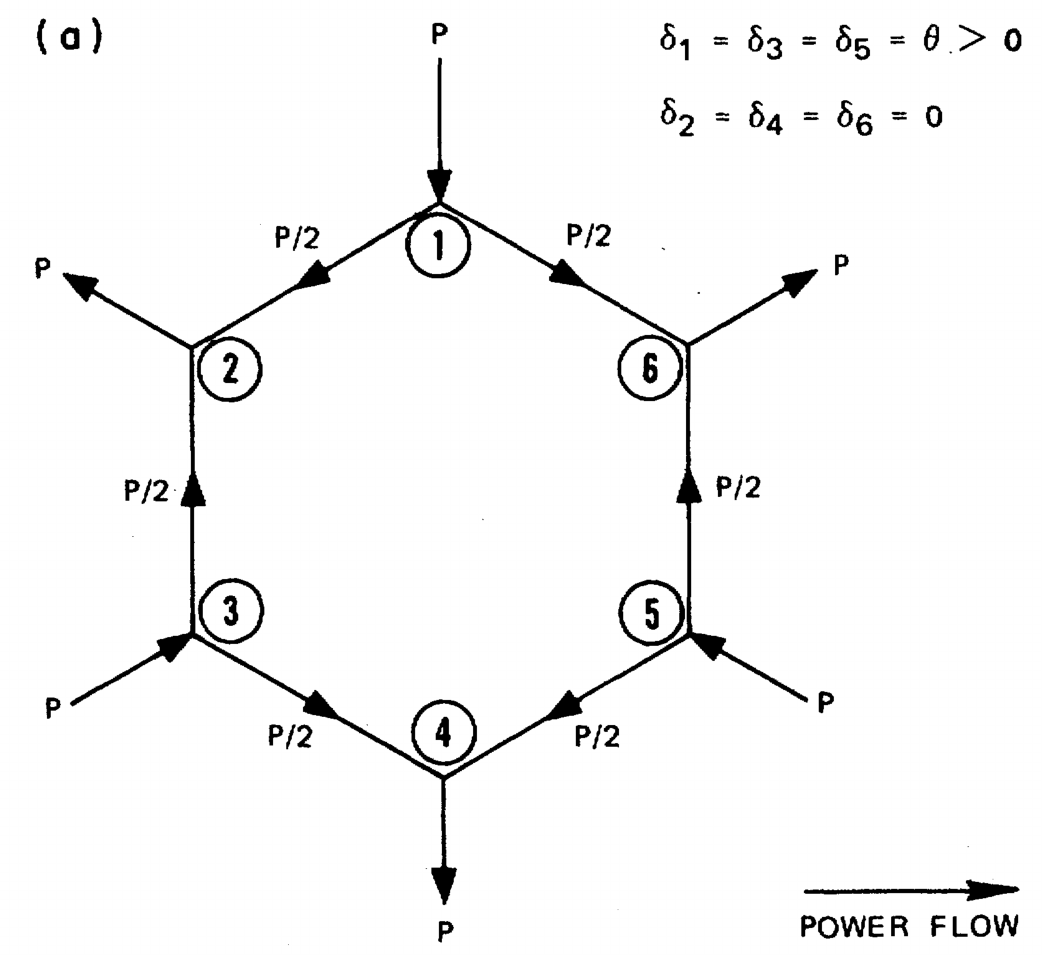}
		\caption{Nominal power flow}
		\label{fig:Korsak1}
	\end{subfigure}
	\begin{subfigure}{0.45\textwidth}
		\includegraphics[width=\textwidth]{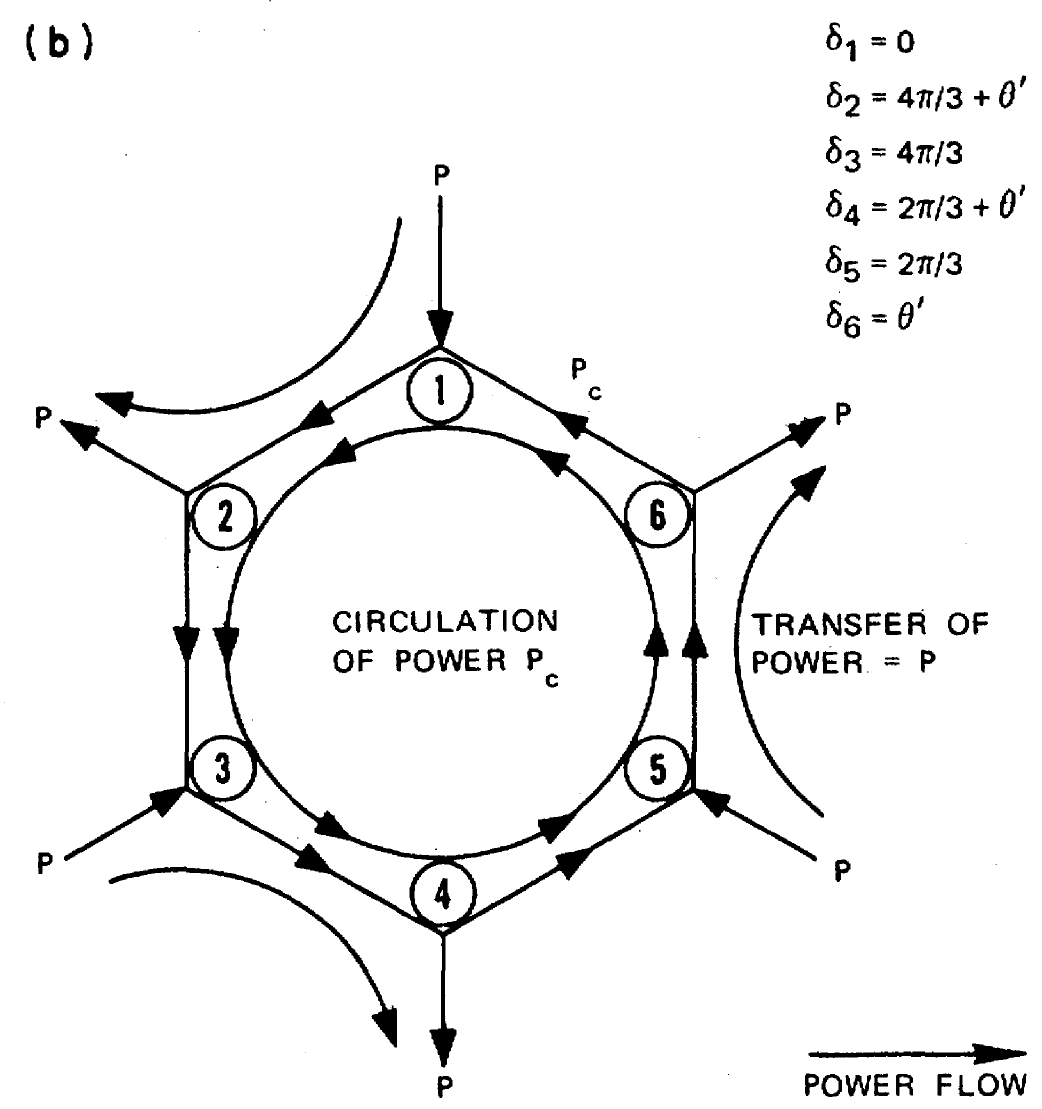}
		\caption{Local power flow}
		\label{fig:Korsak2}
	\end{subfigure}
	\caption{Example of two stable load flow solutions \cite{Korsak}}
	\label{fig:Korsak}
\end{figure}

Network topology of Fig.\ref{fig:Korsak} can have three stable power flow configurations, corresponding to the sum of rotor angles around the loop being $0, 2\pi$ or $-2\pi$. More generally, the maximum number of possible local minima (stable power flows) $N$ depends on the number of generators around the loop $n$ as: $N = 2 \ \text{Int}[n/4] + 1$. Respective rotor angle configurations could sum up around the loop as follows: $\sum_{}^{loop} \delta_{ij} = \pm 2 \pi k \ \text{where} \ k \in \{-a ... 0 ... a\} \ | \ 2a+1=N$ \cite{Korsak}. From equations above one can see the reason of at least 5 generators. Indeed, with more nodes it is possible to have angle loop-sum of an integer multiples of $2\pi$, yet relative angle difference between neighbors is less than $\pi/2$, thus allowing local stability without slipping off.

Obviously, circulating power flows are undesired due to excessive power losses in transmission lines. To deal with this issue, one way to proceed, provided communication between graph neighbors, would be to introduce the notion of communication graph. Note that the method to be proposed relies on real-time angle communication between neighbors. A decentralized and more robust approach could be possible, however, it requires more thorough study/understanding of circulating power emergence, which was not main objective of this work.

\section{Angle consensus in a cycled graph}
Identical angle consensus was already reported in \cite{EnergySyn}, which proposes energy-based control for network flow stabilization in high-order multi-machine power system. In short, key idea in that work, referred to as an incremental energy function approach, was to come up with (state- and communicated angle-dependent) energy function gradient, acting on a machine torque balance equation as a closed-loop input, which, based on Lyapunov-LaSalle argument, stabilizes the whole system at identical angle configuration. Inspired by proposed approach, this section addresses the question of identical angle consensus in cycled graphs based on reduced model.

Arguments of \cite{EnergySyn}, in general, were independent of graph topology and, therefore, are applicable to cycled graphs too. Nevertheless, a separate study of consensus in cycled graphs through a reduced model is still attractive due to aforementioned issue of cycled topologies.

Now, consider the following scenario: triangular network (Fig.\ref{fig:cycled-graph}), reduced model (Fig.\ref{fig:reduced-model}) with objective of rotor-angle consensus (all rotor angles are equal, so no power flow in symmetric network). Reduced model dynamic equations in $\alpha\beta$-frame look like:
\begin{equation}
\label{eq: reduced-system-ab}
\begin{split}
\dot\theta &= \omega
\\
M\dot\omega &= -D\omega + \tau_m - I_r^*(L_m \otimes e_2^\top) \mathbf{R}_\theta^\top \mathbf{E} i_\mathsf{t} 
\\
L_\mathsf{t}\dot i_\mathsf{t} &= -R_\mathsf{t} i_\mathsf{t} + \mathbf{E}^\top \mathbf{R}_\theta (L_m \otimes e_2) I_r^* \omega
\end{split}
\end{equation}

Explanations of various terms in (\ref{eq: reduced-system-ab}) are summarized in Table \ref{t1: terms-explained}:

\begin{table}[h]
	\centering
	\caption{Summary of model-related variables}
	\label{t1: terms-explained}
	\begin{tabular}{ | P{3.5cm} || P{2cm} | P{9cm} | }
		\hline
		Variable & Dimension & Explanation\\
		\hline\hline
		$\theta$ & $\mathbb{T}^n$ & rotor angles of $n$ generators\\
		$\omega$ & $\real^n $ & generator angular frequencies\\
		$M, D$ & $\real^{n \times n}$ & gen. rotational inertia \& damping diag. matrices\\
		$\tau_m$ & $\real^n $ & mechanical torque inputs\\
		$I_r^*$ & $\real^n$ & rotor current set-points \\
		$L_m$ & $\real^n$ & mutual coupling inductances  \\
		$\E$ & $\real^{2n \times 2m}$ & expanded incidence matrix $E \otimes I_2$\\
		$\Lt, R_\mathsf{t}$ & $\real^{2m \times 2m}$ & expanded line parameter matrices $L_t,R_t \otimes I_2$ \\
		$i_\mathsf{t}$ & $\real^{2m}$ & transmission line currents in $\alpha\beta$-frame\\
		$I_r^* (L_m \otimes e_2^\top) \mathbf{R}_\theta^\top \E i_\mathsf{t}$ & $\real^n$ & electrical torques \\
		$\mathbf{R}_\theta (L_m \otimes e_2) I_r^* \omega$ & $\real^{2n}$ & electro-motive forces in $\alpha\beta$-frame  \\
		\hline
	\end{tabular}
\end{table}

The same model in $\mathsf{dq}$-frame looks like:
\begin{equation}
\label{eq: reduced-system-dq}
\begin{split}
\dot\theta_\mathsf{dq} &= \tilde{\omega}
\\
M\dot{\tilde{\omega}} &= -D\tilde{\omega} + \tilde{\tau}_m - I_r^*(L_m \otimes e_2^\top) \mathbf{R}_{\theta_\mathsf{dq}}^\top \mathbf{E} i_\mathsf{t,dq} 
\\
L_\mathsf{t}\dot i_\mathsf{t,dq} &= -\mathbf{Z}_\mathsf{t} i_\mathsf{t,dq} + \mathbf{E}^\top \mathbf{R}_{\theta_\mathsf{dq}} (L_m \otimes e_2) I_r^* (\tilde{\omega} + \omega_0 \boldsymbol{1})
\end{split}
\end{equation}
Here $\omega = \tilde{\omega} + \omega_0 \boldsymbol{1}$, $\tau_m = \tilde{\tau}_m + D\omega_0$, $\tilde{\tau}_m$ is a mechanical torque input in $\mathsf{dq}$-frame, while $D\omega_0$ is a steady-state torque. Rising the question of appropriate $\tilde{\tau}_m$, in general, it is desired to design this closed-loop input such that possible local minima of energy function in cycled graph case are eliminated. Graph topology affects rotor dynamics through electrical torque vector, therefore incremental energy function approach, used in \cite{EnergySyn}, cannot be directly applied. Indeed, mentioned approach uses line current measurements just to modify electrical torque, or, in other words, to arrive at the gradient of aforementioned problematic energy function. Modifying electrical torque in some other way could potentially solve the problem, but, having the design freedom of our closed-loop input and angle communication, it seems more wise to eliminate the effect of cycled graph totally by canceling electrical torque and adding the desired gradient on top. The gradient of interest should not have local minima problems in its associated energy function, just as if the graph did not have any cycles at all. 

\begin{figure}[h]
	\centering
	\includegraphics[height=5cm]{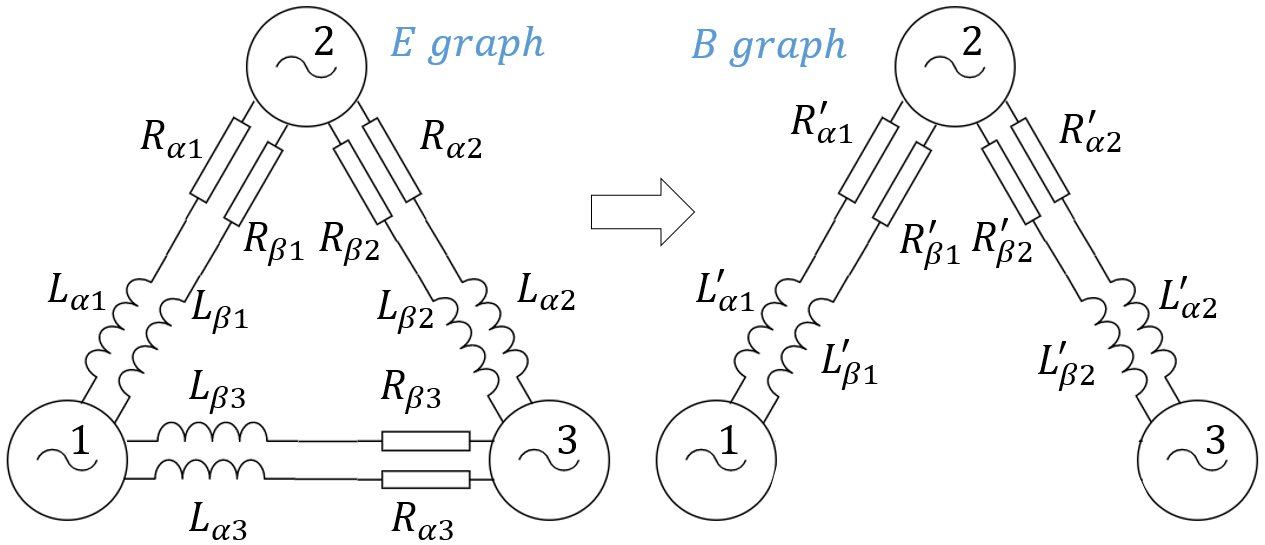}
	\caption{Connection graph $\E$ and communication graph $\mathbf{B}$}
	\label{fig:E-and-B-graphs}
\end{figure}

Bearing in mind the considerations above, one proposes to define a separate communication graph $\mathbf{B}$ that connects all the nodes of $\E$ graph and has all the desired properties, such as no cycles, appropriate edge weights, topology etc. Angles are to be communicated based on $\mathbf{B}$ rather than $\E$ graph neighborhood, gradient term of the torque is to be created as if real connection is described by $\mathbf{B}$ and not by $\E$ graph. Example modification of the graph in Fig.\ref{fig:cycled-graph} is depicted in Fig.\ref{fig:E-and-B-graphs}.

So, with the notion of communication graph we kind of "deceive" our generators and make them think that they are in an un-cycled graph with incidence matrix $\mathbf{B}$ rather than in cycled graph $\E$. Summarizing the closed-loop input properties stated before, one suggests the following candidate: $\tilde{\tau}_m = \tau_e - \nabla_{\theta_\mathsf{dq}} S_B$. Closed-loop dynamics with such feedback become:
\begin{equation}
\label{eq: reduced-system-dq-closed-loop}
\begin{split}
\dot\theta_\mathsf{dq} &= \tilde{\omega}
\\
M\dot{\tilde{\omega}} &= -D\tilde{\omega} - \nabla_{\theta_\mathsf{dq}} S_B
\\
L_\mathsf{t}\dot i_\mathsf{t,dq} &= -\mathbf{Z}_\mathsf{t} i_\mathsf{t,dq} + \mathbf{E}^\top \mathbf{R}_{\theta_\mathsf{dq}} (L_m \otimes e_2) I_r^* (\tilde{\omega} + \omega_0 \boldsymbol{1})
\end{split}
\end{equation} 
Line and rotor dynamics are now decoupled (no dependence on $i_\mathsf{t}$ of the first two equations of (\ref{eq: reduced-system-dq-closed-loop})) with rotor driving line dynamics. This approach resembles feedback linearization. Associated energy function can be defined as $S_B = \frac{1}{2} \Psi_{\mathsf{dq}}^\top BKB^\top \Psi_{\mathsf{dq}}$, where $\Psi_{\mathsf{dq}} = \mathbf{R}_{\theta_\mathsf{dq}} (L_m \otimes e_1) I_r^* \boldsymbol{1}$, $BKB^\top$ - new un-cycled communication graph with edge weight matrix $K$, that has only one stable diagonal on $\mathbb{T}^n$ (property of un-cycled graphs). The gradient can be derived as
\begin{equation}
\label{eq: gradient-Sb}
\nabla_{\theta_\mathsf{dq}} S_B = I_r^* (L_m \otimes e_2^\top) \mathbf{R}_{\theta_\mathsf{dq}}^\top BKB^\top \mathbf{R}_{\theta_\mathsf{dq}} (L_m \otimes e_1) I_r^* \boldsymbol{1}
\end{equation} 
The convergence set of rotor dynamics is $M_1 = \{(\theta_\mathsf{dq},\omega):\omega=\omega_0\boldsymbol{1}, \nabla_{\theta_\mathsf{dq}} S_B = 0\}$. This set is a main diagonal in $\mathbb{T}^n$ (state space of $\theta_\mathsf{dq}$). Line currents, being a simple LTI system under constant input entering in form of steady state angle $\theta_{\mathsf{dq}}^*$, converge to $i_\mathsf{t,dq}^* = \mathbf{Z}_\mathsf{t}^{-1} \mathbf{E}^\top \mathbf{R}_{\theta_\mathsf{dq}^*} (L_m \otimes e_2) I_r^* \omega_0 \boldsymbol{1}$, so that entire system state converges to $M = M_1 \times \{i_\mathsf{t,dq}:i_\mathsf{t,dq} = i_\mathsf{t,dq}^*\}$. Moreover, in case $L_m I_r^* \in \real^{n \times n}$ has same elements along its diagonal, $\mathbf{E}^\top \mathbf{R}_{\theta_\mathsf{dq}^*} (L_m \otimes e_2) I_r^* \omega_0 \boldsymbol{1} = 0 \Rightarrow i_\mathsf{t,dq}^* = 0 \Rightarrow$ no transmission line currents. For this controller one requires line current measurements near every node and rotor angles of neighbors in communication graph. For a simple network of two generators and one transmission line normalized energy function $S_B(\theta_\mathsf{dq})$ can be drawn as in Fig.\ref{fig:2-node-energy-function}:

\begin{figure}[h]
	\centering
	\includegraphics[width=0.79\linewidth]{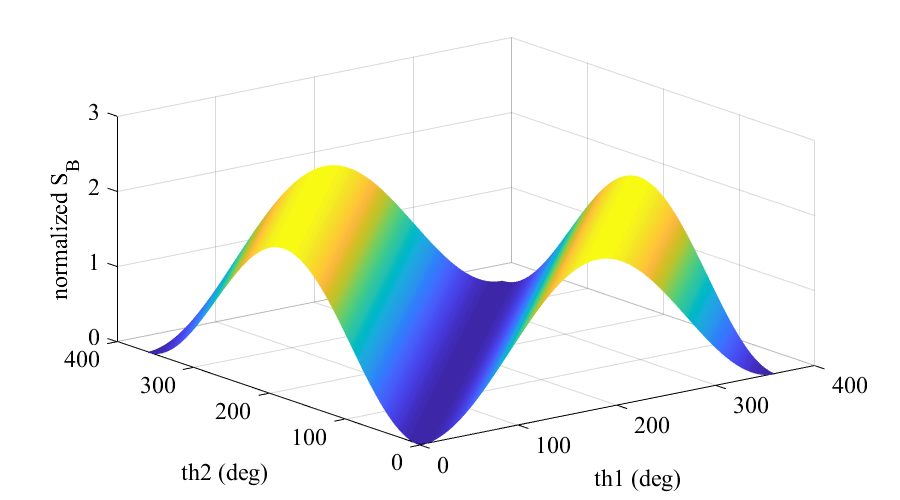}
	\caption{Two-node gradient-creating normalized energy function $S_B$}
	\label{fig:2-node-energy-function}
\end{figure}

\begin{figure}[h]
	\centering
	\begin{subfigure}{0.49\linewidth}
		\includegraphics[width=\linewidth]{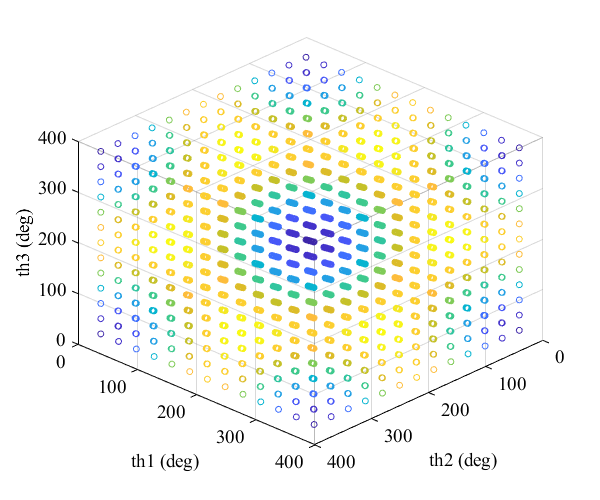}
		\caption{3-node normalized $S_E$}
		\label{fig:3-node-SE}
	\end{subfigure}
	\begin{subfigure}{0.49\linewidth}
		\includegraphics[width=\linewidth]{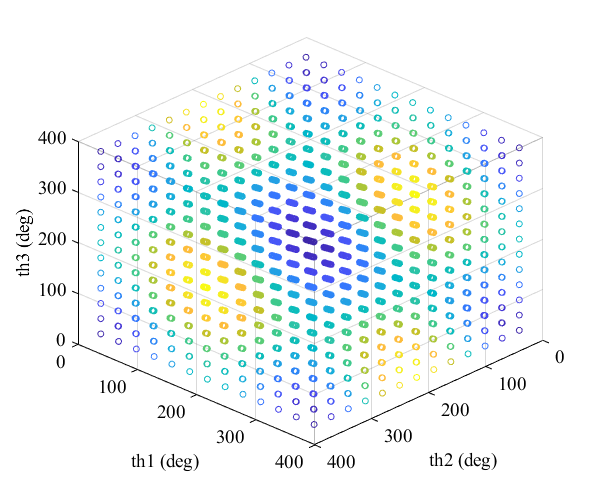}
		\caption{3-node normalized $S_B$}
		\label{fig:3-node-SB}
	\end{subfigure}
	\caption{Three-node gradient-creating normalized energy functions}
	\label{fig:3-node-energy-function}
\end{figure}

For three-node case, depicted in Fig.\ref{fig:E-and-B-graphs}, normalized energy function can still be drawn, but in that case all three axes depict rotor angles, while normalized energy value is conveyed through color. So called "potential" plot for both connection $\E$ and communication $\mathbf{B}$ graphs is shown in Fig.\ref{fig:3-node-energy-function}.

As it can be seen from Fig.\ref{fig:3-node-SE} no local non-diagonal minima are predicted for 3-node cycled graph, as expected. Nevertheless, comparing Fig.\ref{fig:3-node-SE} and Fig.\ref{fig:3-node-SB} interesting observation can be made. In both cases angle consensus (main diagonal at which all angles are equal) is almost globally asymptotically stable, but in $S_B$ case gradient is not as steep (which can be seen from less-bright regions around diagonal) as for $S_E$ case, potentially resulting in longer and larger transients.

\section{Shifted angle consensus in a cycled graph}
Having discussed identical angle consensus, this section focuses on a more interesting case of shifted angle consensus, corresponding to non-zero transmission line currents and, consequently, power transfer between graph nodes. The general question of interest could be formulated as follows: given the desired angle difference in $\mathsf{dq}$-frame, how to design the control input such that mentioned angle configuration is achieved? Note that only angle differences and not their exact values in $\mathsf{dq}$-frame are of interest because transmission line currents and associated power transfer depend on the difference of applied node EMFs. Exact angle values affect only the phase shift of voltage and current vectors back in $\alpha\beta$-frame, as was discussed at the end of Chapter 2.

To solve the issue of shifted consensus, desired angle configuration should be somehow incorporated into the gradient-creating energy function. At least these three energy-function-candidates can be considered due to non-Euclidean nature of angle variable:

\begin{subequations}
	\label{eq: energy-function-candidates}
	\begin{align}
	\bar{S}_B &= S_B(\theta_\mathsf{dq}-\theta_\mathsf{dq}^*) = \frac{1}{2} \Psi^\top (\theta_\mathsf{dq}-\theta_\mathsf{dq}^*) BKB^\top \Psi (\theta_\mathsf{dq}-\theta_\mathsf{dq}^*) \label{eq: SBar}
	\\
	\tilde{S}_B &= \frac{1}{2} \big(\Psi^\top (\theta_\mathsf{dq}) - \Psi^\top (\theta_\mathsf{dq}^*)\big) BKB^\top \big(\Psi (\theta_\mathsf{dq}) - \Psi (\theta_\mathsf{dq}^*)\big) \label{eq: STilde}
	\\
	\hat{S}_B &= S_B(\theta_\mathsf{dq}) - S_B(\theta_\mathsf{dq}^*) - \nabla_{\theta_\mathsf{dq}}^\top S_B(\theta_\mathsf{dq}^*) (\theta_\mathsf{dq}-\theta_\mathsf{dq}^*) \label{eq: SHat}
	\end{align}	
\end{subequations} 

Here $\Psi (\theta_\mathsf{dq}-\theta_\mathsf{dq}^*) = \mathbf{R}_{(\theta_\mathsf{dq} - \theta_\mathsf{dq}^*)} (L_m \otimes e_1) I_r^* \boldsymbol{1}$ and $\Psi_{\mathsf{dq}}^*= \Psi(\theta_\mathsf{dq}^*) = \mathbf{R}_{\theta_\mathsf{dq}^*} (L_m \otimes e_1) I_r^* \boldsymbol{1}$. Equation (\ref{eq: SHat}) is referred to as a Bregman energy function \cite{Bregman}. The Matlab plots of these energies for various angle differences are depicted below:

\begin{figure}[h]
	\centering
	\includegraphics[width=1\textwidth]{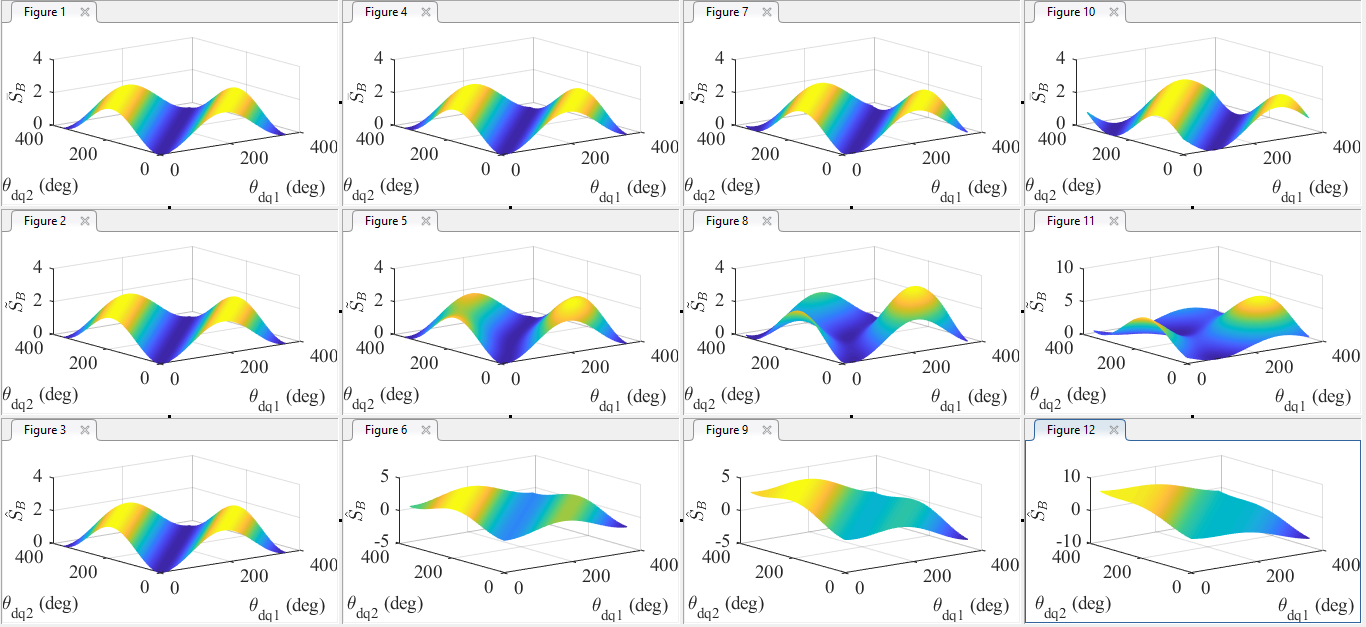}
	\caption{Energy-function-candidate plots for various angle differences}
	\label{fig:enery-function-candidate-plots}
\end{figure}

As one can see, there are 12 realizations with every row corresponding to respective energy function in equation (\ref{eq: energy-function-candidates}), while every column being an increasing angle difference: $(0,0), (0,10), (0,30), (0,90)$. The first energy function nicely stabilizes the shifted diagonal in angle space (xy-plane in this case). Second energy function also has some angles stabilized, but the emergence of several valleys implies that invariant set is disjount, meaning that depending on initial condition, one could end up in wrong valley (angle configuration) and, consequently, wrong power flow. Third candidate, called Bregman energy function turns out to be discontinuous in its $0\leftrightarrow2\pi$ transition, meaning that it is not a good choice for our LaSalle-based stability proof, which requires continuity \cite{Bullo}. All in all, provided communication, the best way (among considered ones) to stabilize the shifted angle consensus in a cycled graph of SG/inverters is to apply the proposed closed-loop input with the gradient of the first energy function.

Regarding the interpretation of proposed candidates, first of all, remember that we are talking about virtual energy functions - fictitious virtually implemented functions that aim to stabilize prescribed configuration in angle space at steady state. They may not always have real energy counterparts, nevertheless, by looking at their structure and noticing the structural similarities with some real energy functions, yet to be utilized in decentralization chapter, a comprehensive insight can be gained on the meaning of those candidates.

The first one, $\bar{S}_B$ (\ref{eq: SBar}), is the "energy" of angle difference. Its idea is to minimize the difference between $\theta_{ij}$ and $\theta_{ij}^*$ in angle space. Since angle difference directly in angle space and not in the Euclidean embedding is achievable by multiplication of respective rotation matrices $\mathbf{R}_{\theta_{\mathsf{dq}}}$ and $\mathbf{R}_{\theta_{\mathsf{dq}}^*}$, the resulting function looks like a somehow bended manifold in Euclidean space - a result of multiplication by desired angle rotations $\mathbf{R}_{\theta_{\mathsf{dq}}^*}$ and, consequently, does not seem to have a real energy counterpart. From matrix perspective, Laplacian of communication graph $\mathbf{B}$ in equation (\ref{eq: SBar}) provides necessary permutation and scaling of sine and cosine terms that, after application of trigonometric rules, results in components having the structure such as $sin(\theta_{ij}-\theta_{ij}^*)$, which is almost globally stable at desired angle consensus $\theta_{ij}=\theta_{ij}^*$.

The second candidate, $\tilde{S}_B$ (\ref{eq: STilde}), is the "energy" of flux linkage difference. Its idea is to minimize the difference between flux linkages $\Psi(\theta_\mathsf{dq}) \ \text{and} \ \Psi(\theta_\mathsf{dq}^*)$ in Euclidean space. From structural perspective transmission line energy stored in inductances $\Lt$ is very similar, which will become evident in simplified model decentralized control section of next chapter, so $\tilde{S}_B$ can be given the interpretation of transmission line energy. Moreover, proposed interpretation explains the emergence of disjoint invariant set in Fig.\ref{fig:enery-function-candidate-plots}. It is known that transmission line currents (and therefore energy) depend on node voltage differences rather than their absolute value. Undesired valley in angle space occurs due to constant step up in reference voltages throughout the graph. This effect is kind of circulating current counterpart in voltage notation, since it adds to node voltages, but not to line currents. From matrix perspective, valley is explained by non-zero kernel of incidence matrix. Indeed, the difference of angle embeddings in Euclidean space could have a constant shift multiplied by vector of all ones ($\boldsymbol{1}$), which lies in kernel of incidence matrix and therefore will not be observed neither in energy function nor in its gradient.

Regarding Bregman energy function, $\hat{S}_B$ (\ref{eq: SHat}), its virtue is the capability of shifting the gradient itself, rather than the references of its entries in some form. This candidate was not studied extensively after observing aforementioned discontinuity, thus exploring this option still remains an interesting open topic for further work.

\section{Remarks on full model, robustness and decentralization}

Some brief remarks, observations and further considerations conclude Chapter 3:\\

$\bullet$ \ Regarding full model, control strategy does not change. The only difference is in driven system, which has stator (or output filter in inverter notation) in addition to line dynamics. Driven system still goes to its own invariant set in angle-frequency state space, while enlarged driven LTI system goes to its own steady state under effect of constant angle input, entering as $\mathbf{R}_{\theta_{\mathsf{dq}}^*}$.\\

$\bullet$ \ As it was mentioned, gradient creation requires angle communication, which is not a robust implementation in real time. As a decentralized alternative, a fully decoupled energy function below could be considered:

\begin{equation}
\label{eq:decoupled-energy-function}
S_B = \frac{1}{2} (\Psi_{\mathsf{dq}} - \Psi_{\mathsf{dq}}^*)^\top (L_m \otimes I_2)^{-1} (\Psi_{\mathsf{dq}} - \Psi_{\mathsf{dq}}^*)
\end{equation} 

where $\Psi_{\mathsf{dq}}^* = \mathbf{R}_{\theta_\mathsf{dq}^*} (L_m \otimes e_1) I_r^* \boldsymbol{1}$. With this as an input, generators do not care about the network and simply converge to their angle references,  stabilizing only a single point and not the whole diagonal in angle space, potentially resulting in longer and larger transients.\\

$\bullet$ \ Although what was proposed in (\ref{eq:decoupled-energy-function}) is a decentralized realization, which eliminates the necessity of communication altogether, it is not robust at least with respect to controller clock drift. If one allows a tiny drift in clocks of neighboring controllers, as time evolves the relative angle and, consequently, the whole network, will drift away from its OPF set-points with controllers being not even aware of the problem, because what they do after all is compare self-measured (or -created in case of inverter) angle with their references, both evolving using same local clock. Of course one way to solve the communication problem would be to synchronize clocks and continuie the work in that direction, but ideally it is not a good practice to fully decouple subsystems. Some kind of feedback is still desired to preclude the adverse effects of inevitable parameter fluctuations and model inaccuracies.\\

Robustness issues and coupling arguments above motivate studying "coupled" decentralization, which is the culminating chapter of the thesis.

\chapter{Decentralization}

This chapter is dedicated to the decentralized control framework of small-cycled or un-cycled multi-machine power systems. For now large-cycled graphs were eliminated from considerations to ensure no sticking into local minimum corresponding to circulating power flows during transients is possible. The work is conducted using both reduced and full models interchangeably and the chapter is structured in chronological order. First, several unsuccessful trials of designing decentralized controller for full models are shown with respective issues discussed. Second, after facing difficulties in tackling full model directly, decentralized controller for reduced model, proposed recently by authors of \cite{EnergySyn}, is reviewed in search of inspiration. Third, led by an idea of simplifying the full model to the level of reduced one, from where a modified controller of previous section could be applied, a simplified full model decentralized controller and associated simplification conditions are discussed. It will be evident that there are several major problems associated with simplification approach, which motivate studying full model back again. Finally, inspired by last section's controller as well as topology-independence remarks, the chapter culminates with a "coupled" decentralized controller for full model. Numerical simulations, advantages and challenges of the proposed controller are discussed in the end. Note that due to excessive amount of formulas, several obvious steps are omitted for the sake of readability and simplicity.

\section{Full model approaches}

\subsection{Feedback linearization}

This approach was already discussed in a concluding remark of topology-independence chapter. Below are some useful reminder matrix equations as well as full model dynamics under one-way-coupled (or feedback-linearized) input:\\
$[\hat{i}_{s,\mathsf{dq}}^\top, \hat{v}_{\mathsf{dq}}^\top, \hat{i}_{\mathsf{t,dq}}^\top]^\top = \boldsymbol{\Pi} \xi_\mathsf{dq}$ \ \ $[\hat{i}_{s}^\top, \hat{v}^\top, \hat{i}_{\mathsf{t}}^\top]^\top = \boldsymbol{\Pi} \xi$ \ \ $\boldsymbol{\Pi_2} = \big[\Zs(\Yc + \mathcal{L}_\mathsf{t}) + I\big]^{-1}$ \ \
$v_\mathsf{dq}^* = \boldsymbol{\Pi_2} \xi^*$
\begin{equation}
\label{eq: full-model-open-loop-ab}
\begin{split}
\dot\theta &= \omega
\\
M\dot{\omega} &= -D\omega + \tau_m - I_r^* (L_m \otimes e_2^\top) \mathbf{R}_{\theta}^\top i_{s}
\\
L_s \dot{i}_{s} &= -R_s i_{s} + \Rtheta (L_m \otimes e_2) I_r^* \omega - v
\\
C \dot{v} &= i_s - \E i_\mathsf{t} - G v
\\
\Lt \dot{i}_\mathsf{t} &= - R_\mathsf{t} i_\mathsf{t} + \Et v
\end{split}
\end{equation}

Same equation in $\mathsf{dq}$-frame and error coordinates for driving system looks like:

\begin{equation}
\label{eq: full-model-open-loop-dq}
\begin{split}
\dot\theta_\mathsf{dq} &= \tilde\omega
\\
M\dot{\tilde\omega} &= -D\tilde\omega +\tilde \tau_m - I_r^* (L_m \otimes e_2^\top) \mathbf{R}_{\theta_\mathsf{dq}}^\top i_{s,\mathsf{dq}}
\\
L_s \dot{i}_{s,\mathsf{dq}} &= -\Zs i_{s,\mathsf{dq}} + \Rtheta (L_m \otimes e_2) I_r^* (\tilde{\omega} + \omega_0 \boldsymbol{1}) - \Vdq
\\
C \dot{v}_\mathsf{dq} &= \is - \E \itdq - \Yc \Vdq
\\
\Lt \itdqdot &= - \Zt \itdq + \Et \Vdq
\end{split}
\end{equation}

Closed-loop input comes from Section 3.2: $\tilde{\tau}_m = \tau_e - \nabla_{\theta_\mathsf{dq}} S_B = I_r^* (L_m \otimes e_2^\top) \mathbf{R}_{\theta_\mathsf{dq}}^\top i_{s,\mathsf{dq}} - \nabla_{\theta_\mathsf{dq}} S_B$, where one could use real $\E$ or communication $\mathbf{B}$ graph. Remaining arguments are as before: driving system convergence to its own almost globally asymptotically stable invariant set, driven system, containing now stator dynamics and still being an LTI system under constant input, converges to angle-dictated steady state. The problem with this approach, namely real-time angle communication, still persists.

\subsection{Incremental energy function}
Define closed-loop input: $\tilde{\tau}_m = I_r^* (L_m \otimes e_2^\top) \Rtheta^\top (K i_{s,\mathsf{dq}} + i_{s,\mathsf{dq}}^{*im})$, where $K = \mathbf{Re}(\Pi_1) \cdot \Pi_1^{-1} = \Pi^\top \begin{bmatrix}
R_s & & \\ & G & \\ & & R_\mathsf{t}
\end{bmatrix} \Pi \cdot \Pi_1^{-1}$ and $i_{s,\mathsf{dq}}^{*im} = \mathbf{Im}(\Pi_1) \cdot \xi^* = \Pi^\top \begin{bmatrix}
\boldsymbol{j} \omega_0 L_s & & \\ & -\boldsymbol{j} \omega_0 C & \\ & & \boldsymbol{j} \omega_0 L_\mathsf{t}
\end{bmatrix} \Pi \cdot \xi^*$. Closed-loop dynamics look like:

\begin{equation} 
\label{eq: two-way-one-closed-loop-dq}
\begin{split}
\dot\theta_\mathsf{dq} &= \tilde\omega
\\
M\dot{\tilde\omega} &= -D\tilde\omega - I_r^* (L_m \otimes e_2^\top) \mathbf{R}_{\theta_\mathsf{dq}}^\top \big(I-K\big) \tilde{i}_{s,\mathsf{dq}} - \underbrace{I_r^* (L_m \otimes e_2^\top) \mathbf{R}_{\theta_\mathsf{dq}}^\top \big(\hat{i}_{s,\mathsf{dq}}^{im} - i_{s,\mathsf{dq}}^{*im}\big)}_\text{$\nabla S$}
\\
L_s \dot{\tilde{i}}_{s,\mathsf{dq}} &= -\Zs \tilde{i}_{s,\mathsf{dq}} + \Rtheta (L_m \otimes e_2) I_r^* \tilde{\omega} - \tilde{v}_\mathsf{dq} - \omega_0 \boldsymbol{j} L_s \Pi_1 \Rtheta (L_m \otimes e_2) I_r^* \tilde{\omega}
\\
C \dot{\tilde{v}}_\mathsf{dq} &= \tilde{i}_{s,\mathsf{dq}} - \E \tilde{i}_{\mathsf{t,dq}} - \Yc \tilde{v}_\mathsf{dq} - \omega_0 \boldsymbol{j} C \Pi_2 \Rtheta (L_m \otimes e_2) I_r^* \tilde{\omega}
\\
\Lt \dot{\tilde{i}}_{\mathsf{t,dq}} &= - \Zt \tilde{i}_{\mathsf{t,dq}} + \Et \tilde{v}_\mathsf{dq} - \omega_0 \boldsymbol{j} \Lt \Pi_3 \Rtheta (L_m \otimes e_2) I_r^* \tilde{\omega}
\end{split}
\end{equation}

where $S = \frac{1}{2} \big(\hat{\xi} - \xi^*\big)^\top \Pi^\top \begin{bmatrix}
L_s & & \\ & -C & \\ & & L_\mathsf{t}
\end{bmatrix} \Pi \big(\hat{\xi} - \xi^*\big)$ can be seen as a Lagrangian function. Consider now the following full-system energy equation:
\begin{equation}
\label{eq: two-way-one-energy-function}
\tilde H = \frac{1}{2} \tilde\omega^\top M \tilde\omega + \frac{1}{2} \tilde{i}_{s,\mathsf{dq}}^\top L_s \tilde{i}_{s,\mathsf{dq}} + \frac{1}{2} \tilde{v}_\mathsf{dq}^\top C \tilde{v}_\mathsf{dq} + \frac{1}{2} \tilde{i}_\mathsf{t,dq}^\top L_\mathsf{t} \tilde{i}_\mathsf{t,dq} + S(\theta_\mathsf{dq})
\end{equation}
Expression for its Lie derivative looks like:
\begin{equation} 
\begin{split}
\label{eq:two-way-one-lie-derivative}
&\dt \tilde H = \begin{bmatrix}
\tilde{\omega} \\ \tilde{i}_{s,\mathsf{dq}} \\ \tilde{v}_\mathsf{dq} \\ \tilde{i}_\mathsf{t,dq}
\end{bmatrix}^\top
\underbrace{\begin{bmatrix}
	-D & & & \\
	\Big[\big(\omega_0 \boldsymbol{j} L_s\big)^\top \Pi_1 + K^\top \Big]\mathbf{R}_{\theta_\mathsf{dq}} (L_m \otimes e_2) I_r^* & -R_s & & \\
	\big(\omega_0 \boldsymbol{j} C\big)^\top \Pi_2 \mathbf{R}_{\theta_\mathsf{dq}} (L_m \otimes e_2) I_r^* & & -G & \\
	\big(\omega_0 \boldsymbol{j} \Lt\big)^\top \Pi_3 \mathbf{R}_{\theta_\mathsf{dq}} (L_m \otimes e_2) I_r^* & & & -R_\mathsf{t}\\
	\end{bmatrix}}_\text{-Q}
\begin{bmatrix}
\tilde{\omega} \\ \tilde{i}_{s,\mathsf{dq}} \\ \tilde{v}_\mathsf{dq} \\ \tilde{i}_\mathsf{t,dq}
\end{bmatrix} < 0.
\end{split}
\end{equation}

From equation (\ref{eq:two-way-one-lie-derivative}) one could derive explicitly the inequality condition on damping term $D$ such that almost global asymptotic stability is achieved. Note that whenever angle space penetrates system state-space (which is the case for both reduced and full model), almost global asymptotic stability can be achieved at most because the angle space itself is closed and bounded, meaning that it is inherently impossible to have a radially-unbounded energy function with level sets projected onto the state-space. Nonetheless, maximum energy points with zero gradient (hills of energy plot) are unstable solutions, so there is no need to worry about those in practice.

The problem with defined closed-loop input is that matrix $K$ is not block-diagonal, meaning that it is impossible to create such input using local stator current measurements only. Difficulties in tackling full model directly motivate reviewing reduced model decentralized control.

\section{Reduced model decentralized control}
This section reviews the decentralized controller proposed recently by authors of \cite{EnergySyn}. Reduced model open-loop dynamics in $\mathsf{dq}$-frame are the same as in (\ref{eq: reduced-system-dq}):
\begin{equation}
\label{eq: reduced-system-dq2}
\begin{split}
\dot\theta_\mathsf{dq} &= \tilde{\omega}
\\
M\dot{\tilde{\omega}} &= -D\tilde{\omega} + \tilde{\tau}_m - I_r^*(L_m \otimes e_2^\top) \mathbf{R}_{\theta_\mathsf{dq}}^\top \mathbf{E} i_\mathsf{t,dq} 
\\
L_\mathsf{t}\dot i_\mathsf{t,dq} &= -\mathbf{Z}_\mathsf{t} i_\mathsf{t,dq} + \mathbf{E}^\top \mathbf{R}_{\theta_\mathsf{dq}} (L_m \otimes e_2) I_r^* (\tilde{\omega} + \omega_0 \boldsymbol{1})
\end{split}
\end{equation}
Proposing the following closed-loop input in $\mathsf{dq}$-frame: $\tilde{\tau}_m = I_r^* (L_m \otimes e_2^\top) \mathbf{R}_{\theta_\mathsf{dq}}^\top \E \mathbf{Z}_\mathsf{t}^{-\top} R_\mathsf{t} \itdq$, one could arrive at the following closed-loop system dynamics in $\mathsf{dq}$:
\begin{equation}
\label{eq: reduced-system-dq-closed-loop2}
\begin{split}
\dot\theta_\mathsf{dq} &= \tilde{\omega}
\\
M\dot{\tilde{\omega}} &= -D\tilde{\omega} + I_r^* (L_m \otimes e_2^\top) \mathbf{R}_{\theta_\mathsf{dq}}^\top \E \mathbf{Z}_\mathsf{t}^{-\top} \boldsymbol{j} \omega_0 \Lt \tilde{i}_\mathsf{t,dq} \\&+ I_r^* (L_m \otimes e_2^\top) \mathbf{R}_{\theta_\mathsf{dq}}^\top \E \mathbf{Z}_\mathsf{t}^{-\top} \boldsymbol{j} \omega_0 \Lt \mathbf{Z}_\mathsf{t}^{-1} \Et \Rtheta (L_m \otimes e_2) I_r^* \omega_0 \boldsymbol{1} = 
\\
&= -D\tilde{\omega} + I_r^* (L_m \otimes e_2^\top) \mathbf{R}_{\theta_\mathsf{dq}}^\top \E \mathbf{Z}_\mathsf{t}^{-\top} \boldsymbol{j} \omega_0 \Lt \tilde{i}_\mathsf{t,dq} - \nabla_{\theta_\mathsf{dq}} S
\\
L_\mathsf{t} \dot{\tilde{i}}_\mathsf{t,dq} &= -\mathbf{Z}_\mathsf{t} \tilde{i}_\mathsf{t,dq} + R_\mathsf{t} \mathbf{Z}_\mathsf{t}^{-1} \mathbf{E}^\top \mathbf{R}_{\theta_\mathsf{dq}} (L_m \otimes e_2) I_r^* \tilde{\omega}
\end{split}
\end{equation} 
Here $\nabla_{\theta_\mathsf{dq}} S = \frac{\partial^\top}{\partial \theta_\mathsf{dq}} (\frac{1}{2} \hat{i}_\mathsf{t,dq}^\top L_\mathsf{t} \hat{i}_\mathsf{t,dq})$. Now, due to assumption of $\frac{\Lt}{R_\mathsf{t}} = const$ the proposed closed-loop input is decentralized. Indeed, thanks to aforementioned assumption, a constant block-diagonal matrix $\mathbf{Z}_\mathsf{t}^{-\top} \boldsymbol{j} \omega_0 \Lt$ now commutes with incidence matrix $\E$ in (\ref{eq: reduced-system-dq-closed-loop2}), which is quite useful, since the resulting $\E \itdq$ is nothing else that just a vector of current injections into the network, readily available for measurements locally. The closed-loop dynamics above are almost globally asymptotically stable, which can be proven by LaSalle argument in the following manner:
\begin{equation}
\label{eq: reduced-model-energy-function1}
\tilde H = \frac{1}{2} \tilde\omega^\top M \tilde\omega + \frac{1}{2} \tilde{i}_\mathsf{t,dq}^\top L_\mathsf{t} \tilde{i}_\mathsf{t,dq} + S(\theta_\mathsf{dq})
\end{equation}
Expression for proposed energy function's Lie derivative looks like:
\begin{equation}
\begin{split}
\label{eq: reduced-model-lie-derivative1}
&\dt \tilde H = -\tilde\omega^\top D \tilde\omega - \tilde{\omega}^\top \nabla_{\theta_\mathsf{dq}} S + \tilde{\omega}^\top I_r^* (L_m \otimes e_2^\top) \mathbf{R}_{\theta_\mathsf{dq}}^\top \E \mathbf{Z}_\mathsf{t}^{-\top} \boldsymbol{j} \omega_0 \Lt \tilde{i}_\mathsf{t,dq} - \tilde{i}_\mathsf{t,dq}^\top R_\mathsf{t} \tilde{i}_\mathsf{t,dq} + \\
&+ \tilde{i}_\mathsf{t,dq}^\top R_\mathsf{t} \mathbf{Z}_\mathsf{t}^{-1} \Et \mathbf{R}_{\theta_\mathsf{dq}} (L_m \otimes e_2) I_r^* \tilde{\omega} + (\nabla_{\theta_\mathsf{dq}} S)^\top \tilde{\omega} = \\
&= \begin{bmatrix}
\tilde{\omega} \\ \tilde{i}_\mathsf{t,dq}
\end{bmatrix}^\top
\underbrace{\begin{bmatrix}
	-D & I_r^* (L_m \otimes e_2^\top) \mathbf{R}_{\theta_\mathsf{dq}}^\top \mathbf{E} \mathbf{Z}_\mathsf{t}^{-\top} \boldsymbol{j} \omega_0 L_\mathsf{t} \\
	R_\mathsf{t} \mathbf{Z}_\mathsf{t}^{-1} \Et \mathbf{R}_{\theta_\mathsf{dq}} (L_m \otimes e_2) I_r^* & -R_\mathsf{t}
	\end{bmatrix}}_\text{-Q}
\begin{bmatrix}
\tilde{\omega} \\ \tilde{i}_\mathsf{t,dq}
\end{bmatrix} < 0
\end{split}
\end{equation}
From here, based on $(Q+Q^\top)/2 \succ 0$ condition, the explicit inequality for damping $D$ that ensures almost global asymptotic stability and should be satisfied for all $\theta_\mathsf{dq}$, can be derived:
\begin{equation}
\label{eq: reduced-model-explicit-D-condition1}
D \succ I_r^* (L_m \otimes e_2^\top) \mathbf{R}_{\theta_\mathsf{dq}}^\top \mathbf{E} \frac{R_\mathsf{t}^{-1}}{4} \Et \Rtheta (L_m \otimes e_2) I_r^*
\end{equation}

Proposed controller is of an incremental energy function-type. It stabilizes the consensus in angle space. To shift the desired references an additional locally-measurable term should be added to the closed-loop input: $\tilde{\tau}_m = I_r^* (L_m \otimes e_2^\top) \mathbf{R}_{\theta_\mathsf{dq}}^\top \E \mathbf{Z}_\mathsf{t}^{-\top} R_\mathsf{t} \itdq - I_r^* (L_m \otimes e_2^\top) \mathbf{R}_{\theta_\mathsf{dq}}^\top \E \mathbf{Z}_\mathsf{t}^{-\top} \boldsymbol{j} \omega_0 \Lt \mathbf{Z}_\mathsf{t}^{-1} \Et \mathbf{R}_{\theta_{\mathsf{dq}}^*} (L_m \otimes e_2) I_r^* \omega_0 \boldsymbol{1}$. Now the shifted gradient and associated energy function would look like: $\nabla_{\theta_\mathsf{dq}} S = \frac{\partial^\top}{\partial \theta_\mathsf{dq}} (\frac{1}{2} (\hat{i}_\mathsf{t,dq}-i_\mathsf{t,dq}^*)^\top L_\mathsf{t} (\hat{i}_\mathsf{t,dq}-i_\mathsf{t,dq}^*))$. Since this energy function does not enter the equation on full energy derivative (\ref{eq: reduced-model-lie-derivative1}), inequality condition on damping term $D$ remains unchanged.

Note that $S(\theta_{\mathsf{dq}})$ has the structure of second energy function $\tilde{S}_B$ in our topology-independence discussion. From structural similarity one could argue that virtual energy function $\tilde{S}_B$ back in Chapter 3 had the interpretation of transmission line inductive energy.

\section{Simplified full model decentralized control}
The idea behind this section is to simplify the full model to the level of reduced one from where a modified version of last section's controller could be applied. It turns out that simplification is possible under the following three conditions, which will become evident in further analysis:

$\bullet$ \ Static stator assumption or sufficient time scale separation ($\tau_{stator} << \tau_{line}$)

$\bullet$ \ $L_\mathsf{t}/R_\mathsf{t} = const$ for transmission lines as in Chapter 2

$\bullet$ \ Block diagonal dominance of $\Pi_2$-map (or $\Pi_1$)

We first review the full model in both stationary and rotating frames:
\begin{equation}
\label{eq: high-order-system}
\begin{cases}
\begin{split}
\dot\theta &= \omega
\\
M\dot\omega &= -D\omega +\tau_m +\tau_e
\\
\dot\lambda_r &= - R_r i_r + u_r
\\
\dot\lambda_s &= - R_s i_s + v
\\
C\dot v &= -Gv - i_s - \mathbf{E}i_\mathsf{t}
\\
L_\mathsf{t}\dot i_\mathsf{t} &= -R_\mathsf{t} i_\mathsf{t} + \mathbf{E}^\top v
\end{split}
\end{cases}
\Leftrightarrow
\begin{cases}
\begin{split}
\dot\theta_\mathsf{dq} &= \tilde\omega
\\
M\dot{\tilde\omega} &= -D\tilde\omega +\tilde \tau_m + \tau_e
\\
\dot\lambda_r &= - R_r i_r + u_r
\\
\dot\lambda_{s,\mathsf{dq}} &= - \mathbf{Z}_s i_{s,\mathsf{dq}} + v_\mathsf{dq} - \omega_0\boldsymbol{j}\Psi_{s,\mathsf{dq}} 
\\
C\dot v_\mathsf{dq} &= -\mathbf{Y}_c v_\mathsf{dq} - i_{s,\mathsf{dq}} - \mathbf{E}i_\mathsf{t,dq}
\\
L_\mathsf{t}\dot i_\mathsf{t,dq} &= -\mathbf{Z}_\mathsf{t} i_\mathsf{t,dq} + \mathbf{E}^\top v_\mathsf{dq}
\end{split}
\end{cases}
\end{equation}
Here we have used the following notation for angle embedding $\Psi_{s,\mathsf{dq}} =  \mathbf{R}_{\theta_\mathsf{dq}}(L_m \otimes \mathbf{e}_1)i_r$. Under assumption of static stator, simplified model dynamics in dq-reference frame can be written as: 
\begin{equation}
\label{eq: model-static-stator}
\begin{split}
\dot\theta_\mathsf{dq} &= \tilde\omega
\\
M\dot{\tilde\omega} &= -D\tilde\omega +\tilde \tau_m - I_r^* (L_m \otimes e_2^\top) \mathbf{R}_{\theta_\mathsf{dq}}^\top i_{s,\mathsf{dq}}
\\
\Zs \is &= \Rtheta (L_m \otimes e_2) I_r^* (\tilde{\omega} + \omega_0 \boldsymbol{1}) - \Vdq
\\
\Yc \Vdq &= \is - \E \itdq
\\
\Lt \itdqdot &= - \Zt \itdq + \Et \Vdq
\end{split}
\end{equation}
As it can be seen from equation (\ref{eq: simple-model-static-stator}) static stator assumption eliminates respective inductance and capacitance dynamics, leaving their steady state equations only. Expressing $\is$ and $\Vdq$ in terms of $\itdq$, system (\ref{eq: simple-model-static-stator}) can be reduced to: 
\begin{equation}
\label{eq: simple-model-static-stator}
\begin{split}
\dot\theta_\mathsf{dq} &= \tilde\omega
\\
M\dot{\tilde\omega} &= -D\tilde\omega +\tilde \tau_m - I_r^* (L_m \otimes e_2^\top) \mathbf{R}_{\theta_\mathsf{dq}}^\top (\mathbf{Z}_s + \mathbf{Y}_c^{-1})^{-1} \Rtheta (L_m \otimes e_2) I_r^* (\tilde{\omega} + \omega_0 \boldsymbol{1}) \\&- I_r^* (L_m \otimes e_2^\top) \Rtheta^\top (I + \Yc \Zs)^{-1} \mathbf{E} i_\mathsf{t,dq}
\\
L_\mathsf{t}\dot i_\mathsf{t,dq} &= -\mathbf{Z}_\mathsf{t} i_\mathsf{t,dq} + \mathbf{E}^\top (I + \Yc \Zs)^{-1} \Rtheta (L_m \otimes e_2) I_r^* (\tilde{\omega} + \omega_0 \boldsymbol{1}) - \Et (\Zs^{-1} + \Yc)^{-1} \E \itdq \end{split}
\end{equation}
Recall that mechanical torque input previously was defined as $\tau_m = D\omega_0 + \tilde{\tau}_m$. In reduced static stator model (\ref{eq: simple-model-static-stator-reduced}) a stator-dependent damping term allows redefinition of input as follows: $\tau_m = D'\omega_0 + \tilde{\tau}_m$. New damping factor looks like
 \begin{equation}
 \label{eq: D'-definition}
 D' = D + \mathsf{diag} \bigg[(L_m i_r^*)^2 (G + \Arrowvert Y_c \Arrowvert^2 R_s) \frac{\Arrowvert Y_c \Arrowvert^2}{\Arrowvert A \Arrowvert^2}\bigg]_k
 \end{equation}
where $\Arrowvert Y_c \Arrowvert^2 = (G^2 + \omega_0^2 C^2)$ and $\Arrowvert A \Arrowvert^2 = (G + R_s \Arrowvert Y_c \Arrowvert^2)^2 + \omega_0^2 (C - \Arrowvert Y_c \Arrowvert^2 L_s)^2$ are scalar values corresponding to every separate node. Moreover, transmission line impedance matrix $\Zt$ can also be redefined as: 
 \begin{equation}
\label{eq: Zt'-definition}
\Zt' = \Zt + \Et (\Zs^{-1} + \Yc)^{-1} \E
\end{equation}
With these in mind, reduced model simplifies even further:
\begin{equation}
\label{eq: simple-model-static-stator-reduced}
\begin{split}
\dot\theta_\mathsf{dq} &= \tilde\omega
\\
M\dot{\tilde\omega} &= -D'\tilde\omega +\tilde \tau_m - I_r^* (L_m \otimes e_2^\top) \Rtheta^\top (I + \Yc \Zs)^{-1} \mathbf{E} i_\mathsf{t,dq}
\\
L_\mathsf{t}\dot i_\mathsf{t,dq} &= -\Zt' \itdq + \mathbf{E}^\top (I + \Yc \Zs)^{-1} \Rtheta (L_m \otimes e_2) I_r^* (\tilde{\omega} + \omega_0 \boldsymbol{1})
\end{split}
\end{equation}
As a next step, recall the error-coordinate relation $\tilde{i}_{\mathsf{t,dq}} = i_{\mathsf{t,dq}} - \hat{i}_{\mathsf{t,dq}}(\theta_\mathsf{dq})$ and high-order model steady-state map $\hat{i}_{\mathsf{t,dq}} = \Pi_3 \xi = \mathbf{Z}_\mathsf{t}^{-1} \mathbf{E}^\top \Pi_2 \xi$. Using these, one can derive the derivative of steady-state current $\dot{\hat{i}}_\mathsf{t,dq} = \mathbf{Z}_\mathsf{t}^{-1} \mathbf{E}^\top \Pi_2 \omega_0 \boldsymbol{j} \mathbf{R}_{\theta_\mathsf{dq}} (L_m \otimes \mathbf{e}_2) I_r^* \tilde{\omega}$ and rewrite (\ref{eq: simple-model-static-stator-reduced}) as:
\begin{equation}
\label{eq: simple-model-static-stator-reduced-itopen}
\begin{split}
\dot\theta_\mathsf{dq} &= \tilde\omega
\\
M\dot{\tilde\omega} &= -D'\tilde\omega +\tilde \tau_m - I_r^* (L_m \otimes e_2^\top) \Rtheta^\top (I + \Yc \Zs)^{-1} \mathbf{E} i_\mathsf{t,dq}
\\
L_\mathsf{t}\dot{\tilde{i}}_\mathsf{t,dq} &= -\Zt' \tilde{i}_\mathsf{t,dq} + \big(\mathbf{E}^\top (I + \Yc \Zs)^{-1} - \omega_0 \boldsymbol{j} \Lt \mathbf{Z}_\mathsf{t}^{-1} \mathbf{E}^\top \Pi_2 \big)\Rtheta (L_m \otimes e_2) I_r^* \tilde{\omega}
\end{split}
\end{equation}
where one used steady state equilibrium condition $\Zt' \hat{i}_\mathsf{t,dq} = \mathbf{E}^\top (I + \Yc \Zs)^{-1} \Rtheta (L_m \otimes e_2) I_r^* \omega_0$. To achieve angle consensus in closed-loop, the following modified control law is suggested:
\begin{equation}
\label{eq: dq-input-suggestion}
\tilde\tau_m = I_r^* (L_m \otimes e_2^\top) \Rtheta^\top (I + \Yc \Zs)^{-1} \E \itdq\ +\ I_r^* (L_m \otimes e_2^\top) \mathbf{R}_{\theta_\mathsf{dq}}^\top \Pi_2^\top \mathbf{E} \Zt^{-\top} \omega_0 \boldsymbol{j} \Lt \itdq
\end{equation}
The first term fully cancels the electrical torque, while the second takes care of the negative gradient term. Note that the first term can be generated based on local line current measurements only, thanks to block-diagonal structure of $(I+\Yc\Zs)^{-1}$. The second term' local estimation will be addressed later. Proposed control law allows rewriting (\ref{eq: simple-model-static-stator-reduced-itopen}) in closed-loop form as follows:
\begin{equation}
\label{eq: reduced-model-static-stator-closed-loop}
\begin{split}
\dot\theta_\mathsf{dq} &= \tilde\omega
\\
M\dot{\tilde\omega} &= -D'\tilde\omega + I_r^* (L_m \otimes e_2^\top) \mathbf{R}_{\theta_\mathsf{dq}}^\top \Pi_2^\top \mathbf{E} \mathbf{Z}_\mathsf{t}^{-\top} \omega_0 \boldsymbol{j} L_\mathsf{t} \tilde{i}_\mathsf{t,dq} - \nabla_{\theta_\mathsf{dq}} S
\\
L_\mathsf{t}\dot{\tilde{i}}_\mathsf{t,dq} &= -\Zt' \tilde{i}_\mathsf{t,dq} + \big(\mathbf{E}^\top (I + \Yc \Zs)^{-1} - \omega_0 \boldsymbol{j} \Lt \mathbf{Z}_\mathsf{t}^{-1} \mathbf{E}^\top \Pi_2 \big)\Rtheta (L_m \otimes e_2) I_r^* \tilde{\omega}
\end{split}
\end{equation}
where the gradient term is the same as in previous reduced model controller: 
\begin{equation}
\label{eq: unshifted-gradient}
\nabla_{\theta_\mathsf{dq}} S = -\bar{I}_r^* (L_m \otimes e_2^\top) \bar{\mathbf{R}}_{\theta_\mathsf{dq}}^\top \mathbf{E} \mathbf{Z}_\mathsf{t}^{-\top} \omega_0 \boldsymbol{j} L_\mathsf{t} \mathbf{Z}_\mathsf{t}^{-1} \mathbf{E}^\top \bar{\mathbf{R}}_{\theta_\mathsf{dq}} (L_m \otimes e_2) \bar{I}_r^* \omega_0 \mbf{1} = \frac{\partial^\top}{\partial \theta_\mathsf{dq}} (\frac{1}{2} \hat{i}_\mathsf{t,dq}^\top L_\mathsf{t} \hat{i}_\mathsf{t,dq})
\end{equation}
A subtle difference between previous section's controller and present one is in stator presence. Indeed, if in previous model node electro-motive forces $\xi = \Rtheta (L_m \otimes e_2) I_r^* \omega_0 \mbf{1}$ influenced transmission lines directly, now they do it through stator, which has a steady-state map of $\Pi_2$: $\xi \rightarrow \Vdq$. Therefore, effective rotation map $\bar{\mathbf{R}}_{\theta_\mathsf{dq}}$ and rotor currents $\bar{I}_r^*$, which correspond to $\Vdq$, must be used in gradient. These effective quantities can be interpreted as the amplitudes and dq-angles of node voltages $\Vdq$ throughout the graph. Corresponding analytical relation is: $\Pi_2 \xi = \Pi_2 \Rtheta (L_m \otimes e_2) I_r^* \omega_0 \mbf{1} = \bar{\mathbf{R}}_{\theta_\mathsf{dq}} (L_m \otimes e_2) \bar{I}_r^* \omega_0 \mbf{1} = \Vdq$. 

With this, local generation of the second term in closed-loop input (\ref{eq: dq-input-suggestion}) can be addressed. Notice $\hat{i}_\mathsf{t,dq}^\top/\omega_0 = \mbf{1}^\top I_r^* (L_m \otimes e_2^\top) \mathbf{R}_{\theta_\mathsf{dq}}^\top \Pi_2^\top \mathbf{E} \Zt^{-\top} = \mbf{1}^\top \bar{I}_r^* (L_m \otimes e_2^\top) \bar{\mathbf{R}}_{\theta_\mathsf{dq}}^\top \E \Zt^{-\top} =\mbf{1}^\top \mathsf{diag}^\top(\Vdq/\omega_0) \E \Zt^{-\top}$ with $\hat{i}_\mathsf{t,dq} \in \real^{2n}$ and $\mathsf{diag}(\Vdq/\omega_0) \in \real^{2n \times n}$, meaning that second term in input can be approximated as: $\mathsf{diag}^\top(\Vdq/\omega_0) \E \Zt^{-\top} \omega_0 \boldsymbol{j} \Lt \itdq$ under diagonal dominance of $\Pi_2$. Thus, it also can be generated locally with node voltage and transmission line current measurements, thanks to assumption of constant $R_\mathsf{t}/\Lt$ ratio, which allows commuting $\E$ and $\Zt^{-\top} \omega_0 \boldsymbol{j} \Lt$, and special structure of $\Pi_2$, allowing block-diagonal form of $\mathsf{diag}(\Vdq/\omega_0)$. 
   
Consider now the following full energy function:
\begin{equation}
\label{eq: energy-function}
\tilde H = \frac{1}{2} \tilde\omega^\top M \tilde\omega + \frac{1}{2} \tilde{i}_\mathsf{t,dq}^\top L_\mathsf{t} \tilde{i}_\mathsf{t,dq} + S(\theta_\mathsf{dq})
\end{equation}
Expression for its Lie derivative looks like:
\begin{equation}
\begin{split}
\label{eq: energy-function-lie-derivative}
&\dt \tilde H = -\tilde\omega^\top D' \tilde\omega - \tilde{\omega}^\top \nabla_{\theta_\mathsf{dq}} S + \tilde{\omega}^\top I_r^* (L_m \otimes e_2^\top) \mathbf{R}_{\theta_\mathsf{dq}}^\top \Pi_2^\top \mathbf{E} \mathbf{Z}_\mathsf{t}^{-\top} \boldsymbol{j} \omega_0 L_\mathsf{t} \tilde{i}_\mathsf{t,dq} - \\
&- \tilde{i}_\mathsf{t,dq}^\top \Zt' \tilde{i}_\mathsf{t,dq} + \tilde{i}_\mathsf{t,dq}^\top \big(\Et (I + \Yc\Zs)^{-1} - \omega_0 \boldsymbol{j} L_\mathsf{t} \mathbf{Z}_\mathsf{t}^{-1} \mathbf{E}^\top \Pi_2 \big) \mathbf{R}_{\theta_\mathsf{dq}} (L_m \otimes e_2) I_r^* \tilde{\omega} + (\nabla_{\theta_\mathsf{dq}} S)^\top \tilde{\omega} = \\
&= \begin{bmatrix}
\tilde{\omega} \\ \tilde{i}_\mathsf{t,dq}
\end{bmatrix}^\top
\underbrace{\begin{bmatrix}
-D' & 2I_r^* (L_m \otimes e_2^\top) \mathbf{R}_{\theta_\mathsf{dq}}^\top \Pi_2^\top \mathbf{E} \mathbf{Z}_\mathsf{t}^{-\top} \boldsymbol{j} \omega_0 L_\mathsf{t} \\
\Et (I + \Yc\Zs)^{-1} \mathbf{R}_{\theta_\mathsf{dq}} (L_m \otimes e_2) I_r^* & -\Zt'
\end{bmatrix}}_\text{-Q}
\begin{bmatrix}
\tilde{\omega} \\ \tilde{i}_\mathsf{t,dq}
\end{bmatrix} < 0
\end{split}
\end{equation}
For $\dt \tilde H$ to be negative definite we require that $A = \frac{1}{2} (Q + Q^\top)$ be positive definite, which, recalling equation (\ref{eq: Zt'-definition}), is equivalent to:
\begin{equation}
\label{eq: condition-on-D'}
\begin{split}
D' \succ I_r^* (L_m \otimes e_2^\top) \mathbf{R}_{\theta_\mathsf{dq}}^\top &\Big(\Pi_2^\top \Zt^{-\top} \boldsymbol{j} \omega_0 \Lt + \frac{1}{2} (I + \Yc \Zs)^{-\top} \Big) \E \Big(\Zt + \Et (\Zs^{-1} + \Yc)^{-1} \E \Big)^{-1} \Et \\ &\Big(\frac{1}{2} (I + \Yc\Zs)^{-1} + \omega_0 \boldsymbol{j}^\top \Lt \Zt^{-1} \Pi_2 \Big) \Rtheta (L_m \otimes e_2) I_r^*
\end{split}
\end{equation}
or, using the relation between $D$ and $D'$ (\ref{eq: D'-definition}), equivalent to
\begin{equation}
\begin{split}
\label{eq: condition-on-D}
D &\succ I_r^* (L_m \otimes e_2^\top) \mathbf{R}_{\theta_\mathsf{dq}}^\top \Big(\Pi_2^\top \Zt^{-\top} \boldsymbol{j} \omega_0 \Lt + \frac{1}{2} (I + \Yc \Zs)^{-\top} \Big) \E \Big(\Zt + \Et (\Zs^{-1} + \Yc)^{-1} \E \Big)^{-1} \Et \\ &\Big(\frac{1}{2} (I + \Yc\Zs)^{-1} + \omega_0 \boldsymbol{j}^\top \Lt \Zt^{-1} \Pi_2 \Big) \Rtheta (L_m \otimes e_2) I_r^* - \mathsf{diag} \bigg[(L_m i_r^*)^2 (G + \Arrowvert Y_c \Arrowvert^2 R_s) \frac{\Arrowvert Y_c \Arrowvert^2}{\Arrowvert A \Arrowvert^2}\bigg]_k
\end{split}
\end{equation}
Note that in equations (\ref{eq: condition-on-D'}-\ref{eq: condition-on-D}) $\E$ and $\Zt^{-\top} \boldsymbol{j} \omega_0 \Lt$ commute due to $L_\mathsf{t}/R_\mathsf{t} = \alpha$ assumption. Moreover, the following relation holds: $\mathbf{Z}_\mathsf{t}^{-\top} \omega_0^2 L_\mathsf{t}^2 R_\mathsf{t}^{-1} \mathbf{Z}_\mathsf{t}^{-1} = \frac{\omega_0^2}{\omega_0^2 + \alpha^2} R_\mathsf{t}^{-1}$.\\
For shifted angle consensus in closed-loop, the following control law is suggested:
\begin{equation}
\begin{split}
\label{eq: dq-input-suggestion-shifted}
\tilde\tau_m &= I_r^* (L_m \otimes e_2^\top) \Rtheta^\top (I + \Yc \Zs)^{-1} \E \itdq + I_r^* (L_m \otimes e_2^\top) \mathbf{R}_{\theta_\mathsf{dq}}^\top \Pi_2^\top \mathbf{E} \Zt^{-\top} \omega_0 \boldsymbol{j} \Lt (\itdq-\itdq^*)\\
&\approx I_r^* (L_m \otimes e_2^\top) \Rtheta^\top (I + \Yc \Zs)^{-1} \E \itdq + \mathsf{diag}^\top(\Vdq/\omega_0) \mathbf{E} \Zt^{-\top} \omega_0 \boldsymbol{j} \Lt (\itdq-\itdq^*) 
\end{split}
\end{equation}

Here one adds an imaginary term depending on desired transmission line currents to complete the gradient of a shifted energy function, just as in case of reduced model's controller. Again, thanks to $L_\mathsf{t}/R_\mathsf{t} = \alpha$ assumption, $\mathbf{Z}^{-\top} \omega_0 \boldsymbol{j} L_\mathsf{t}$ is a constant block-diagonal matrix, which commutes with $\mathbf{E}$, meaning that this additional input term can also be generated using local current and voltage information only. Closed-loop dynamics remain same as in (\ref{eq: reduced-model-static-stator-closed-loop}), where
\begin{equation}
\label{eq: shifted-gradient}
\begin{split}
\nabla_{\theta_\mathsf{dq}} S &= -\bar{I}_r^* (L_m \otimes e_2^\top) \bar{\mathbf{R}}_{\theta_\mathsf{dq}}^\top \mathbf{E} \mathbf{Z}_\mathsf{t}^{-\top} \boldsymbol{j} \omega_0 L_\mathsf{t} \mathbf{Z}_\mathsf{t}^{-1} \mathbf{E}^\top (\bar{\mathbf{R}}_{\theta_\mathsf{dq}} - \bar{\mathbf{R}}_{\theta_\mathsf{dq}^*}) (L_m \otimes e_2) \bar{I}_r^* \omega_0 \mbf{1} \\ &= \frac{\partial^\top}{\partial \theta_\mathsf{dq}} \big(\frac{1}{2} (\hat{i}_\mathsf{t,dq} - {i}_\mathsf{t,dq}^*)^\top L_\mathsf{t} (\hat{i}_\mathsf{t,dq} - {i}_\mathsf{t,dq}^*)\big) 
\end{split}
\end{equation}
Since gradient-dependent terms in Lie derivative (\ref{eq: energy-function-lie-derivative}) cancel, conditions on $D$ and $D'$ remain unchanged. 

As a miscellaneous remark, to ensure inequality condition on damping term $D$ for all possible angle combinations during transients, a proportional controller $K_p$ could be added in $\mathsf{dq}$-frame to artificially rise the damping without increasing steady state loss $D\omega_0$, acting as a rotor friction in case of SG and as a DC-link parasitic resistance plus switching losses in case of inverter.

Several Simulink realizations on a triangular network (Fig.\ref{fig:E-and-B-graphs}) with literature-taken specifications show that under sufficiently high $K_p$ and start near steady state one can stabilize the desired OPF set-points with proposed controller acting on a full model. Time scale separation requires $\tau_{lines}/\tau_{stator}$ ratio to be about 25 for all possible transient angle configurations and as small as 8 for some angles. However, there are many parameters to play with, so these numbers are by no means conclusive. More precise conclusions require more simulations, but the message to be drawn from these scarce figures already is that times scale separation does not need to be large to the extent of static stator approximation in order for simplified controller to stabilize full model.

From a higher perspective, simplified model controller has several problems. First, there are three underlying assumptions for it to work, which may not be practical in reality. Moreover, even if all the assumptions are satisfied, gradient-creating energy function of such form almost globally stabilizes a disjoint invariant set in angle space, just as in case of second energy function in out topology-independence discussion, meaning that depending on initial conditions one could end up in wrong "valley", stabilizing undesired angle configuration and, consequently, wrong power flow. The same problem also emerges in reduced model decentralized controller of previous section, which is not surprising, because both use the same energy function and, in fact, one is a slight modification of another. These unacceptable deficiencies motivate returning back to full model altogether. This "journey" to reduced model decentralized control, however, should not be seen as useless, since it turned out to be an intermediate step in full model's decentralized control design, addressed in the next section.

\section{Full model decentralized control}

In this last section a decentralized controller for a full model is developed. First, we review the dynamic equations in $\mathsf{dq}$-frame and error coordinates for rotor dynamics:

\begin{equation}
\label{eq: full-model-open-loop-dq2}
\begin{split}
\dot\theta_\mathsf{dq} &= \tilde\omega
\\
M\dot{\tilde\omega} &= -D\tilde\omega +\tilde \tau_m - I_r^* (L_m \otimes e_2^\top) \mathbf{R}_{\theta_\mathsf{dq}}^\top i_{s,\mathsf{dq}}
\\
L_s \dot{i}_{s,\mathsf{dq}} &= -\Zs i_{s,\mathsf{dq}} + \Rtheta (L_m \otimes e_2) I_r^* (\tilde{\omega} + \omega_0 \boldsymbol{1}) - \Vdq
\\
C \dot{v}_\mathsf{dq} &= \is - \E \itdq - \Yc \Vdq
\\
\Lt \itdqdot &= - \Zt \itdq + \Et \Vdq
\end{split}
\end{equation}

Regarding closed-loop input, a good choice of gradient-creating energy function would preserve coupling, yet avoid non-diagonal matrices as middle arguments to energy function. Moreover, bijective local measurements (such as stator currents and node voltages, not transmission line currents) should be used, because, firstly, these are the only local information sources and network coupling reflections available, and secondly, bijectivity ensures that incidence matrix $\E$ will not enter the gradient and create disjoint invariant sets in angle space (the problem of reduced and simplified-full model controllers). Here a degree of freedom emerges in the form of choosing between stator currents and node voltages. From practical perspective, bus voltages are more important to control and easier to measure/process, so further work will be continued with those. Another approach would be to create both current and voltage-related gradients. This option, however, was not investigated because of possible growth of invariant set in angle space. Indeed, combined gradient could be zero in two cases: when both its current and voltage components are zero and when they are counteracting each other in some directions. This option should not be forgotten though. A suitable voltage-based candidate fulfilling all of these is given below:

\begin{equation}
\begin{split}
\label{eq:full-dq-input-suggestion-shifted}
\tilde\tau_m &= I_r^* (L_m \otimes e_2^\top) \Rtheta^\top i_{s,\mathsf{dq}} + I_r^* (L_m \otimes e_2^\top) \mathbf{R}_{\theta_\mathsf{dq}}^\top \Pi_2^\top \omega_0 \boldsymbol{j} C (v_\mathsf{dq}-v_\mathsf{dq}^*) \\
&\approx I_r^* (L_m \otimes e_2^\top) \Rtheta^\top i_{s,\mathsf{dq}} + \mathsf{diag}^\top(\Vdq/\omega_0) \omega_0 \boldsymbol{j} C (v_\mathsf{dq}-v_\mathsf{dq}^*)
\end{split}
\end{equation}

Notice that there is no incidence matrix $\E$ in the input, thus no need for constant matrix commuting, thus no need in $R_\mathsf{t}/L_\mathsf{t} = const$. Bearing in mind that now we have dynamic (not static) stator, only a diagonal dominance assumption of $\Pi_2$ persists, which was already used in equation (\ref{eq:full-dq-input-suggestion-shifted}). Closed-loop now looks like:

\begin{equation}
\label{eq: full-model-closed-loop-dq2}
\begin{split}
\dot\theta_\mathsf{dq} &= \tilde\omega
\\
M\dot{\tilde\omega} &= -D\tilde\omega + I_r^* (L_m \otimes e_2^\top) \mathbf{R}_{\theta_\mathsf{dq}}^\top \Pi_2^\top \omega_0 \boldsymbol{j} C \tilde{v}_\mathsf{dq} - \nabla_{\theta_\mathsf{dq}} S
\\
L_s \dot{\tilde{i}}_{s,\mathsf{dq}} &= -\Zs \tilde{i}_{s,\mathsf{dq}} + \Rtheta (L_m \otimes e_2) I_r^* \tilde{\omega} - \tilde{v}_\mathsf{dq} - \omega_0 \boldsymbol{j} L_s \Pi_1 \Rtheta (L_m \otimes e_2) I_r^* \tilde{\omega}
\\
C \dot{\tilde{v}}_\mathsf{dq} &= \tilde{i}_{s,\mathsf{dq}} - \E \tilde{i}_{\mathsf{t,dq}} - \Yc \tilde{v}_\mathsf{dq} - \omega_0 \boldsymbol{j} C \Pi_2 \Rtheta (L_m \otimes e_2) I_r^* \tilde{\omega}
\\
\Lt \dot{\tilde{i}}_{\mathsf{t,dq}} &= - \Zt \tilde{i}_{\mathsf{t,dq}} + \Et \tilde{v}_\mathsf{dq} - \omega_0 \boldsymbol{j} \Lt \Pi_3 \Rtheta (L_m \otimes e_2) I_r^* \tilde{\omega}
\end{split}
\end{equation}

where

\begin{equation}
\label{eq: full-shifted-gradient}
\begin{split}
\nabla_{\theta_\mathsf{dq}} S &= -I_r^* (L_m \otimes e_2^\top) \mathbf{R}_{\theta_\mathsf{dq}}^\top \Pi_2^\top \omega_0 \boldsymbol{j} C \Pi_2 \big(\Rtheta-\mathbf{R}_{\theta_\mathsf{dq}^*}\big) (L_m \otimes e_2) I_r^* \omega_0 \boldsymbol{1} \\ 
\nabla_{\theta_\mathsf{dq}} S &= \frac{\partial^\top}{\partial \theta_\mathsf{dq}} \big(\frac{1}{2} (\hat{v}_\mathsf{dq} - {v}_\mathsf{dq}^*)^\top C (\hat{v}_\mathsf{dq} - {v}_\mathsf{dq}^*)\big) 
\end{split}
\end{equation}
Full energy function can be shown as:
\begin{equation}
\begin{split}
\label{eq: full-energy-function}
\tilde H &= \frac{1}{2} \tilde\omega^\top M \tilde\omega + \frac{1}{2} \tilde{i}_{s,\mathsf{dq}}^\top L_s \tilde{i}_{s,\mathsf{dq}} + \frac{1}{2} \tilde{v}_\mathsf{dq}^\top C \tilde{v}_\mathsf{dq} + \frac{1}{2} \tilde{i}_\mathsf{t,dq}^\top L_\mathsf{t} \tilde{i}_\mathsf{t,dq} + S(\theta_\mathsf{dq}) \\ 
\tilde H &= \frac{1}{2} \tilde\omega^\top M \tilde\omega + \frac{1}{2} \tilde{i}_{s,\mathsf{dq}}^\top L_s \tilde{i}_{s,\mathsf{dq}} + \frac{1}{2} \tilde{v}_\mathsf{dq}^\top C \tilde{v}_\mathsf{dq} + \frac{1}{2} \tilde{i}_\mathsf{t,dq}^\top L_\mathsf{t} \tilde{i}_\mathsf{t,dq} + \frac{1}{2} (\hat{v}_\mathsf{dq} - {v}_\mathsf{dq}^*)^\top C (\hat{v}_\mathsf{dq} - {v}_\mathsf{dq}^*)
\end{split}
\end{equation}

Expression for its Lie derivative is:
\begin{equation}
\begin{split}
\label{eq: full-energy-function-lie-derivative}
\dt \tilde H &= -\tilde\omega^\top D \tilde\omega + \tilde{\omega}^\top I_r^* (L_m \otimes e_2^\top) \mathbf{R}_{\theta_\mathsf{dq}}^\top \Pi_2^\top \omega_0 \boldsymbol{j} C \tilde{v}_\mathsf{dq} - \tilde{\omega}^\top \nabla_{\theta_\mathsf{dq}} S - \tilde{i}_{s,\mathsf{dq}}^\top R_s \tilde{i}_{s,\mathsf{dq}} - \tilde{i}_{s,\mathsf{dq}}^\top \tilde{v}_\mathsf{dq} + \\
&+ \tilde{i}_{s,\mathsf{dq}}^\top \Rtheta (L_m \otimes e_2) I_r^* \tilde{\omega} - \tilde{i}_{s,\mathsf{dq}}^\top \omega_0 \boldsymbol{j} L_s \Pi_1 \Rtheta (L_m \otimes e_2) I_r^* \tilde{\omega} + \tilde{v}_\mathsf{dq}^\top \tilde{i}_{s,\mathsf{dq}} - \tilde{v}_\mathsf{dq}^\top \E \tilde{i}_{\mathsf{t,dq}} - \\
&- \tilde{v}_\mathsf{dq}^\top G \tilde{v}_\mathsf{dq} - \tilde{v}_\mathsf{dq}^\top \omega_0 \boldsymbol{j} C \Pi_2 \Rtheta (L_m \otimes e_2) I_r^* \tilde{\omega} - \tilde{i}_\mathsf{t,dq}^\top R_\mathsf{t} \tilde{i}_\mathsf{t,dq} + \tilde{i}_\mathsf{t,dq}^\top \Et \tilde{v}_\mathsf{dq} - \\
& - \tilde{i}_\mathsf{t,dq}^\top \omega_0 \boldsymbol{j} \Lt \Pi_3 \Rtheta (L_m \otimes e_2) I_r^* \tilde{\omega} + (\nabla_{\theta_\mathsf{dq}} S)^\top \tilde{\omega} =\\
\dt \tilde H &=\begin{bmatrix}
\tilde{\omega} \\ \tilde{i}_{s,\mathsf{dq}} \\ \tilde{v}_\mathsf{dq} \\ \tilde{i}_\mathsf{t,dq}
\end{bmatrix}^\top
\underbrace{\begin{bmatrix}
	-D & & & \\
	-\big[\omega_0 \boldsymbol{j} L_s \Pi_1 - I \big]\mathbf{R}_{\theta_\mathsf{dq}} (L_m \otimes e_2) I_r^* & -R_s & & \\
	-2\omega_0 \boldsymbol{j} C \Pi_2 \mathbf{R}_{\theta_\mathsf{dq}} (L_m \otimes e_2) I_r^* & & -G & \\
	-\omega_0 \boldsymbol{j} \Lt \Pi_3 \mathbf{R}_{\theta_\mathsf{dq}} (L_m \otimes e_2) I_r^* & & & -R_\mathsf{t}\\
	\end{bmatrix}}_\text{-Q}
\begin{bmatrix}
\tilde{\omega} \\ \tilde{i}_{s,\mathsf{dq}} \\ \tilde{v}_\mathsf{dq} \\ \tilde{i}_\mathsf{t,dq}
\end{bmatrix} < 0.
\end{split}
\end{equation}

For $\dt \tilde H$ to be negative definite we require that $A = \frac{1}{2} (Q + Q^\top)$ be positive definite, from where explicit inequality condition on damping term $D$ can be derived. This inequality should hold for any angle combination during transients, thus additional control effort in terms of $K_p$ could be used as before.

Regarding the interpretation, $S(\theta_{\mathsf{dq}})$ has similar structure to the forth (fully decoupled) energy function from topology-independence chapter. Indeed, $S(\theta_{\mathsf{dq}})$ is nothing more than the sum of capacitor energies throughout the network and block-diagonal structure of matrix C ensures decentralization. From energy perspective every local controller cares about its own capacitor only. The problem with fully decoupled energy function back in Chapter 3 was lack of coupling and robustness. In this case, although the energy is fully decoupled, coupling enters not through function structure, but in form of measurements, which possess implicit information on neighbor states.

As it was already discussed, only diagonal dominance of $\Pi_2$ is left as a precondition for this controller. The next step would be to check whether slightly violating this condition degrades the controller performance. To check that, the following scenario was simulated on Simulink: triangular network (as before), full model, literature (real life) taken specifications of power system, arbitrary initial conditions and some OPF-prescribed set-points as a local references (Chapter 2). 

\begin{figure}[h]
	\centering
	\begin{subfigure}{0.49\linewidth}
		\includegraphics[width=\linewidth]{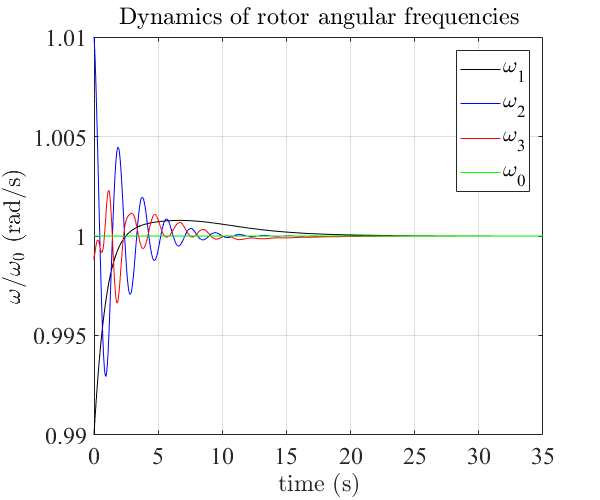}
		\caption{Angular frequency transient}
		\label{fig:omegasP2}
	\end{subfigure}
	\begin{subfigure}{0.49\linewidth}
		\includegraphics[width=\linewidth]{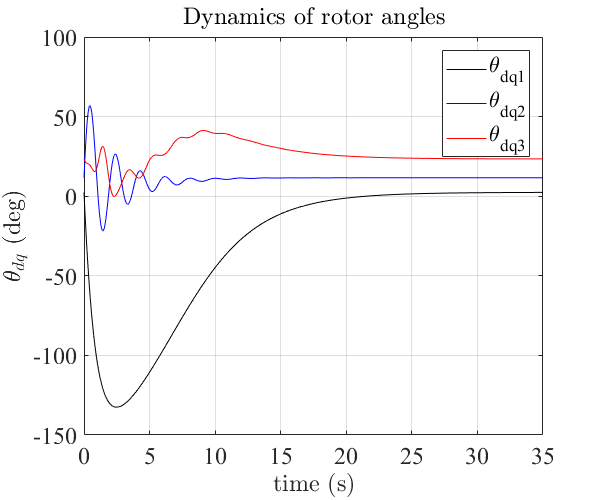}
		\caption{Angles transient}
		\label{fig:anglesP2}
	\end{subfigure}
	\caption{Angular frequency and angle transients with exact controller}
	\label{fig:anglesomegasP2}
\end{figure}

\begin{figure}[h]
	\centering
	\begin{subfigure}{0.49\linewidth}
		\includegraphics[width=\linewidth]{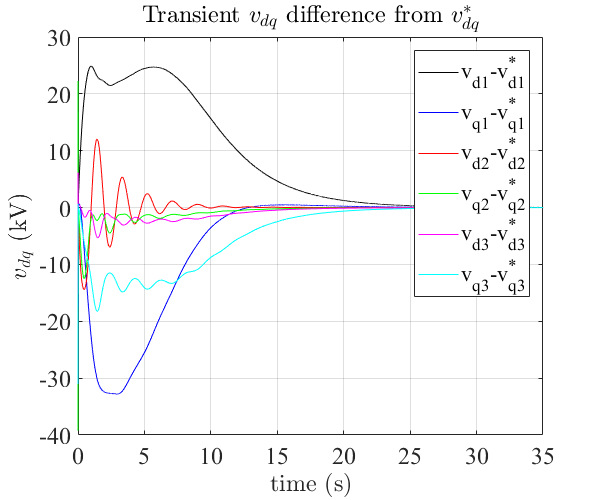}
		\caption{$v_{\mathsf{dq}} - v_{\mathsf{dq}}^*$}
		\label{fig:voltagesP2}
	\end{subfigure}
	\begin{subfigure}{0.49\linewidth}
		\includegraphics[width=\linewidth]{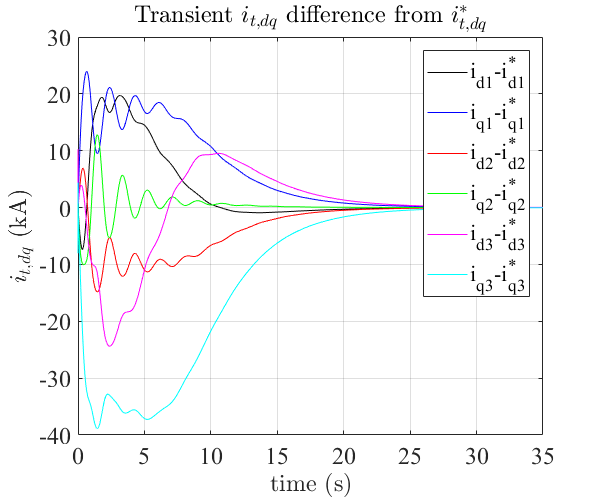}
		\caption{$i_{\mathsf{t,dq}} - i_{\mathsf{t,dq}}^*$}
		\label{fig:currentsP2}
	\end{subfigure}
	\caption{Node voltage and line current transients with exact controller}
	\label{fig:voltagescurrentsP2}
\end{figure}

\begin{figure}[h]
	\centering
	\begin{subfigure}{0.49\linewidth}
		\includegraphics[width=\linewidth]{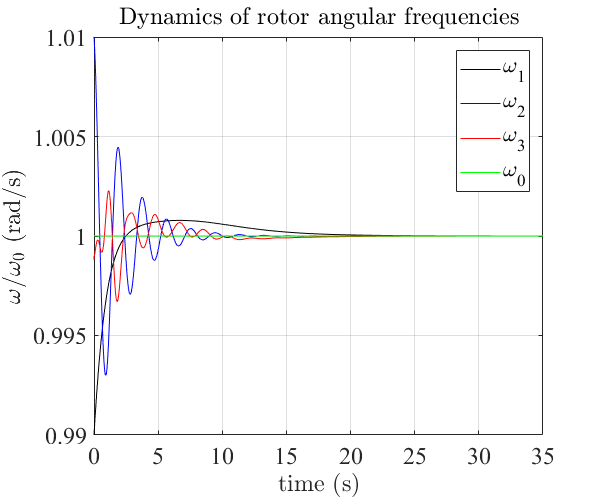}
		\caption{Angular frequency transient}
		\label{fig:omegasVdq}
	\end{subfigure}
	\begin{subfigure}{0.49\linewidth}
		\includegraphics[width=\linewidth]{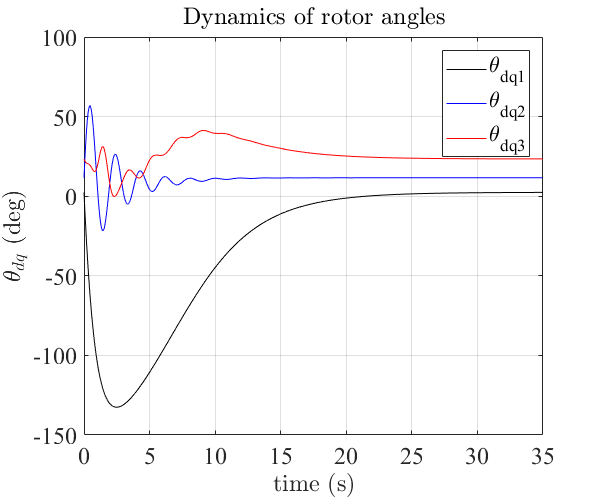}
		\caption{Angles transient}
		\label{fig:anglesVdq}
	\end{subfigure}
	\caption{Angular frequency and angle transients with approximate controller}
	\label{fig:anglesomegasVdq}
\end{figure}

\begin{figure}[h]
	\centering
	\begin{subfigure}{0.49\linewidth}
		\includegraphics[width=\linewidth]{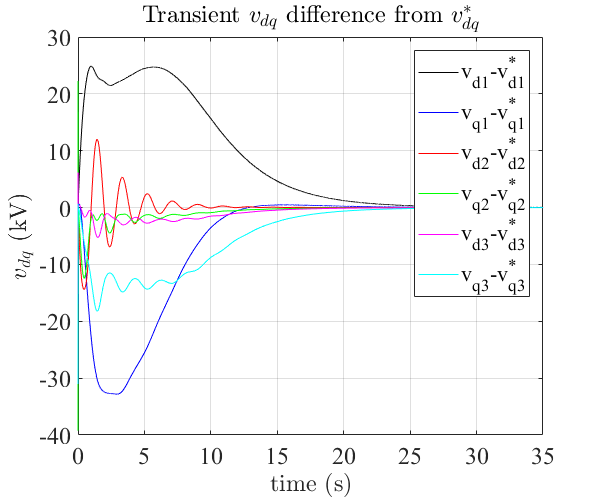}
		\caption{$v_{\mathsf{dq}} - v_{\mathsf{dq}}^*$}
		\label{fig:voltagesVdq}
	\end{subfigure}
	\begin{subfigure}{0.49\linewidth}
		\includegraphics[width=\linewidth]{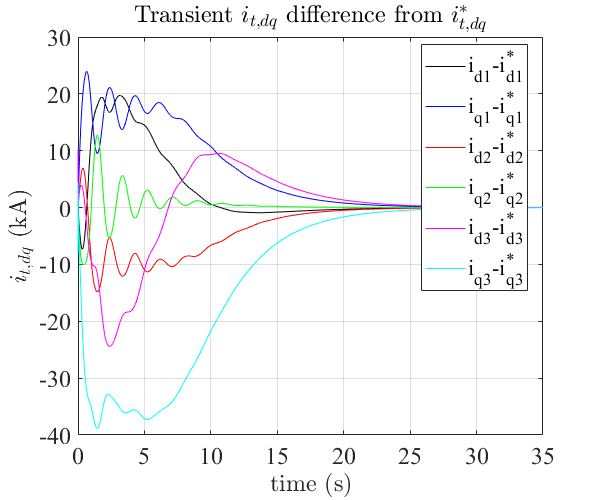}
		\caption{$i_{\mathsf{t,dq}} - i_{\mathsf{t,dq}}^*$}
		\label{fig:currentsVdq}
	\end{subfigure}
	\caption{Node voltage and line current transients with approximate controller}
	\label{fig:voltagescurrentsVdq}
\end{figure}

First, exact controller, given by the first line in equation (\ref{eq:full-dq-input-suggestion-shifted}), was tested. Note that ideally it is not decentralized, but almost global asymptotic stability proof was shown above for it. Rotor angles and angular frequency transients are depicted in Fig.\ref{fig:anglesomegasP2}. As one can see both frequencies and angles converge to the prescribed references and to prove that one also plots the transient difference of node voltages and line currents from their set-points in $\mathsf{dq}$-frame (Fig.\ref{fig:voltagescurrentsP2}).
Second, approximate controller, given by the second line in equation (\ref{eq:full-dq-input-suggestion-shifted}), was tested. This controller is decentralized, but no rigorous stability proof was shown above for it. Rotor angles and angular frequency transients are depicted in Fig.\ref{fig:anglesomegasVdq}. Surprisingly, both frequencies and angles show very similar transients to the ones observed before. To back that up, one also plots the transient difference of node voltages and line currents from their set-points in $\mathsf{dq}$-frame (Fig.\ref{fig:voltagescurrentsVdq}).

\begin{figure}[h]
	\centering
	\includegraphics[width=0.7\linewidth]{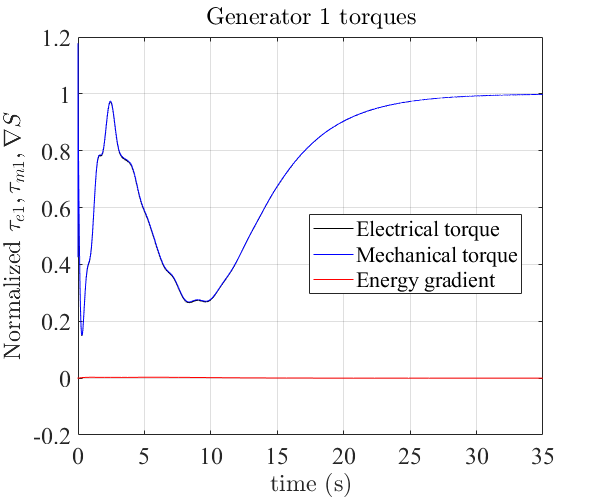}
	\caption{Control action measures for generator 1}
	\label{fig:torques1}
\end{figure}

\begin{figure}[h]
	\centering
	\includegraphics[width=0.7\linewidth]{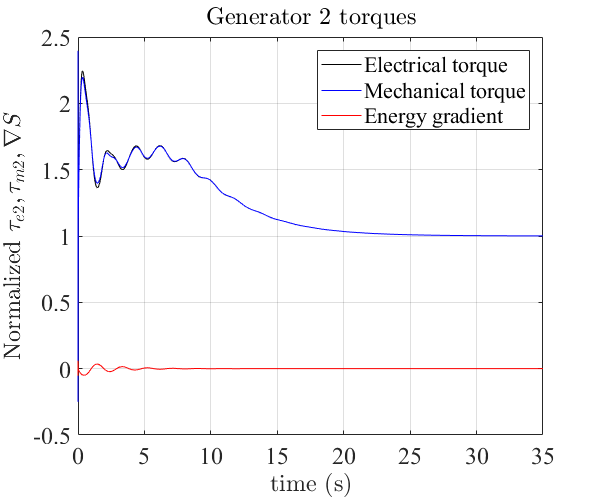}
	\caption{Control action measures for generator 2}
	\label{fig:torques2}
\end{figure}

\begin{figure}[h]
	\centering
	\includegraphics[width=0.7\linewidth]{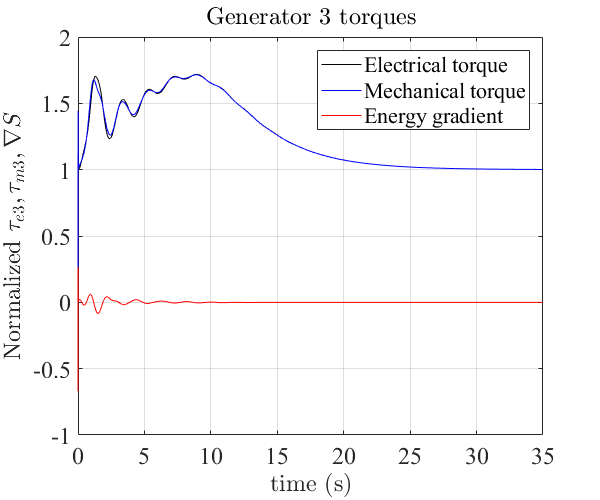}
	\caption{Control action measures for generator 3}
	\label{fig:torques3}
\end{figure}

Regarding the control effort of this decentralized controller, Fig.\ref{fig:torques1}-\ref{fig:torques3} show that major contribution comes from the first term in equation (\ref{eq:full-dq-input-suggestion-shifted}), which deals with canceling electrical torque. Thus, following the torque demand in real time constitutes main control effort, which is not much different from steady state demand as it turns out from transients, shown in Fig.\ref{fig:torques1}-\ref{fig:torques3}.

Overall, it turns out that slight violations of $\Pi_2$'s block-diagonal dominance do not degrade controller performance, which is reassuring, because specifications for simulation were taken from real world MV power system. Those are summarized in Table \ref{t2: parameters}. An in-depth theoretical analysis of $\Pi_2$ internal structure is needed to be on safe side, nevertheless, it can be seen from the first glance that major diagonal dominance violation would require unrealistic parameter combination of a power system. In fact, to preserve diagonal dominance, one requires high local load $G$, high $R_\mathsf{t}$ and $L_\mathsf{t}$ values as well as low $L_s, R_s$. These are compatible with inverter model and voltage level, since relatively low $L_s, R_s$ in combination with moderate $C$ would result in high output filter corner frequency and, consequently, high switching frequency condition, which is exactly the assumption of 3-phase 2-level 6-switch inverter average-switch model within local matching control at low voltage level (LV). Even with MV parameters from Table \ref{t2: parameters}, which do not comply with desired specification description, controller is able to follow the references fairly well as shown by simulations. 

\begin{table}[h]
	\centering
	\caption{Simulated system parameters}
	\label{t2: parameters}
	\begin{tabular}{ | P{3.5cm} || P{3.5cm} | P{7.5cm} | }
		\hline
		Variable & Value & Explanation\\
		\hline\hline
		$\omega_0$ & $100 \pi$ & nominal electrical frequency \\
		$M1,M2,M3$ & $22,10,45 \cdot 10^3$ & Rotor moments of inertia ($kg\cdot m^2$)\\
		$L_{m1},L_{m2},L_{m3}$ & $0.04,0.08,0.02$ & mutual inductances ($H$)\\
		$i_{r1}, i_{r2}, i_{r3}$ & $ 1950,975,3900 $ & reference rotor currents ($A$)\\
		$D_1,D_2,D_3$ & $4.0,1.5,8.5 \cdot 10^3$ & rotor mechanical damping \\				
		$K_{p1},K_{p2},K_{p3}$ & $5D_1,5D_2,5D_3$ & control gain $K_p$ \\
		$G_1,G_2,G_3$ & $0.8,0.4,1.0$ & nominal local loads ($S$) \\
		$R_{\mathsf{t,1}},R_{\mathsf{t,2}},R_{\mathsf{t,3}}$ & $0.165,0.166,0.07$ & line resistances ($\Omega$)\\
		$L_{\mathsf{t,1}},L_{\mathsf{t,2}},L_{\mathsf{t,3}}$ & $4.7,3.8,2.4 \cdot 10^{-3}$ & line inductances ($H$)\\
		$R_{s1},R_{s2},R_{s3}$ & $0.166,0.07,0.5$ & stator resistances ($\Omega$)\\
		$L_{s1},L_{s2},L_{s3}$ & $0.18,0.10,0.66\cdot10^{-3}$ & stator inductances ($H$)\\
		$C_{1},C_{2},C_{3}$ & $0.01,0.2,4\cdot10^{-3}$ & load-side capacitances ($F$)\\
		\hline
	\end{tabular}
\end{table}
\chapter{Conclusion}
\section{Summary}

To sum up, this thesis presents three major results:\\

$\bullet$ \ Bijective translation between network-level OPF-prescribed set-points and local controller references was proposed. This is a major decentralization-enabling step, since decentralized controllers need to have a certificate that converging to local references imposes OPF on the whole network. Bijective translation provides this certificate under certain conditions.\\

$\bullet$ \ Provided communication, circulating power flow problem can be solved using the notion of communication graph. There are associated robustness problems with real-time angle communication and clock drift, which motivate study of decentralization.\\

$\bullet$ \ Motivated by inherent robustness problems associated with communication, a decentralized controller was proposed for full model applied on un-cycled or small-cycled network topologies. Interestingly, proposed controller does not require constant R/L assumption for transmission lines, stator and DC-side time-scale separation and strict diagonal dominance conditions on some matrices, as it turns out from simulations with real-life-taken network parameters.\\

\section{Outlook}

As an outlook, three interesting research directions are indicated:\\

$\bullet$ \ First, it would be interesting to investigate the potential of proposed decentralized controller in dealing with circulating power flows in large cycled topologies. For that, the emergence of circulating power, which was not the focus of this thesis, should be studied more thoroughly.\\

$\bullet$ \ Second, robustness of the controller with respect to load change and/or clock drift should be checked. Throughout the work we considered constant load $G$. Change of parameters in this block-diagonal matrix would affect the $\Pi$-map and, consequently, the available solution manifold in angle-space, meaning that further improvements/changes might be necessary for load-robustness. Regarding clock drift, although decentralized controller realizes the gradient of a decoupled energy function, coupling is still preserved through measurements, utilized in gradient-creation. This "implicit" coupling gives us a hope that the controller can manage clock drift, nevertheless, extensive study/simulations need to be done to have a rigorous proof.\\

$\bullet$ \ Third, mentioned controller should be extended to other types of networks, having only load or generation at some nodes and not both, as it is now. Various types of "non-resistive" loads could be considered. Moreover, inverter average-switch model within local matching control implicitly assumes high switching frequency, meaning low or lowish-medium voltage levels from practical perspective. There is nothing wrong with it, as micro-grid type of systems are most likely to be deployed at residential distribution level, which operates in stated voltage range. However, for densely-populated cities, where there is no much place for distributed generation of renewable energy, it would be useful to step-up voltage level and transmit the power from renewable energy farms. For that purpose, other suitable inverter topologies, such as MMC, could be considered. Benefit and applicability of local model-matching, compatibility with MMC-inherent control objectives, such as voltage-balancing etc., are naturally arising issues in that case.\\

\bibliography{bibliography,long}
\bibliographystyle{IEEEtran}


\section*{Supplementary material}

# ETH

Eidgenössische Technische Hochschule Zürich
Swiss Federal Institute of Technology Zurich

Institut für Automatik

NR.
2018, FS

## SEMESTERARBEIT

IfA-Nr.

| | | |
|---|---|---|
| **Author** | Zhetessov, Aidar | <u>aidarz@student.ethz.ch</u> |
| Office place | ETL I 17.1 | |

**Supervisor**  Catalin Arghir

Issue Date:

Termination Date:

## Virtual-oscillator based control of inverters in power grids: Theory and Experiment

**Description**

In the context of inverter-based power grids we are investigating a control problem formulation starting from average-switch models of 3-phase AC/DC inverters. By neglecting the switching harmonics, the converter system is easier to analyze, retains energy conservation properties and can be still seen as passive. In this setting synchronous machine dynamics can be model-matched, giving rise to a consensus-type problem around the inverters' DC-link voltage.

In this project we consider a multi-inverter scenario connected to resistive loads and study asymptotic stability aspects of angle synchronization under SM-matching control.

**Tasks**

Your task will be to investigate Passivity-based synchronization-control strategies for grid-forming inverters.

**Procedures**

Phase 1: Investigation of various control strategies to incorporate angle set-points and decentralized PLL
Phase 2: Experiments in Simulink and on the Imperix hardware. Several re-iteration to Phase 1 for improvements and generalizations.

**Time schedule**

Phase 1: February - April
Phase 2: April - May

**Oral presentation**  21.05.2018 (tentative)

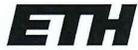
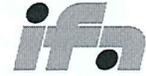

**Eidgenössische Technische Hochschule Zürich**
Swiss Federal Institute of Technology Zurich

**Institut für Automatik**

NR.
2018, FS

# SEMESTERARBEIT

IfA-Nr.

| | | |
|---|---|---|
| **Author** | Zhetessov, Aidar | aidarz@student.ethz.ch |
| Office place | ETL I 17.1 | |

**Signatures**

| | | |
|---|---|---|
| **SUPERVISING PROFESSOR** | Prof. Florian Dörfler, Institut für Automatik | |
| **SUPERVISOR** | Catalin Arghir | |
| **STUDENT** | Aidar Zhetessov | |
2

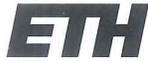

**Eidgenössische Technische Hochschule Zürich**
Swiss Federal Institute of Technology Zurich

## Declaration of originality

The signed declaration of originality is a component of every semester paper, Bachelor's thesis, Master's thesis and any other degree paper undertaken during the course of studies, including the respective electronic versions.

Lecturers may also require a declaration of originality for other written papers compiled for their courses.

I hereby confirm that I am the sole author of the written work here enclosed and that I have compiled it in my own words. Parts excepted are corrections of form and content by the supervisor.

**Title of work** (in block letters):

A DECENTRALIZED CONTROL METHODOLOGY FOR MULTI-MACHINE/MULTI-CONVERTER POWER NETWORKS

**Authored by** (in block letters):
*For papers written by groups the names of all authors are required.*

**Name(s):** ZHETESSOV

**First name(s):** AIDAR

With my signature I confirm that
- I have committed none of the forms of plagiarism described in the 'Citation etiquette' information sheet.
- I have documented all methods, data and processes truthfully.
- I have not manipulated any data.
- I have mentioned all persons who were significant facilitators of the work.

I am aware that the work may be screened electronically for plagiarism.

**Place, date**
ZURICH, 04.06.2018

**Signature(s)**
*[signature]*

*For papers written by groups the names of all authors are required. Their signatures collectively guarantee the entire content of the written paper.*


\end{document}